\documentclass[10pt,aps,preprint,prl,pre,amssymb,twocolumn]{revtex4}

\usepackage{graphicx}
\usepackage{ulem}
\usepackage{subfigure}

\begin{document}

\title{Flame Acceleration in Channels with Obstacles in the Deflagration-to-Detonation Transition}

\author{Damir Valiev$^{1}$}
\author{Vitaly Bychkov$^{1}$}
\author{V'yacheslav Akkerman$^{2}$}
\author{Chung K. Law$^{2}$}
\author{Lars-Erik Eriksson$^{3}$}

\affiliation{$^{1}$Department of Physics, Ume{\aa}  University, 901 87 Ume{\aa}, Sweden \\
$^{2}$Department of Mechanical and Aerospace Engineering, Princeton University, Princeton, NJ 08544-5263, USA \\
$^{3}$Department of Applied Mechanics, Chalmers University of
Technology, 412 96 Gothenburg, Sweden}

\begin{abstract}
It was demonstrated recently in Bychkov et al., Phys. Rev. Lett.
101 (2008) 164501, that the physical mechanism of flame acceleration
in channels with obstacles is qualitatively different from the
classical Shelkin mechanism. The new mechanism is much stronger,
and is independent of the Reynolds number. The present study
provides details of the theory and numerical modeling of the
flame acceleration. It is shown theoretically and computationally
that flame acceleration progresses noticeably faster in the
axisymmetric cylindrical geometry as compared to the planar one,
and that the acceleration rate reduces with increasing initial Mach
number and thereby the gas compressibility. Furthermore, the velocity
of the accelerating flame saturates to a constant value that is
supersonic with respect to the wall. The saturation state can be
correlated to the Chapman-Jouguet deflagration  as well as the fast
flames observed in experiments. The possibility of transition from
deflagration to detonation in the obstructed channels is demonstrated.
\end{abstract}

\maketitle

\section{1. Introduction}
In the process of deflagration-to-detonation transition (DDT) in
tubes with a closed end, a slow premixed flame accelerates
spontaneously from the closed end and triggers detonation
\cite{Shelkin,Zeldovich.et.al-1985,Utriev&Oppenheim-1966,Shepherd.et.al-1992,
Roy-et-al-2004,Ciccarelli-Dorofeev-2008,Ciccarelli-et-al-2005,Kuznetsov-et-al-2005,
Frolov-et-al-2007,Johansen-Ciccarelli-2007,Johansen-Ciccarelli-2009,
Gamezo-et-al-2007,Gamezo-et-al-2008}. A qualitative explanation of
the process was first proposed by Shelkin, Ref. \cite{Shelkin}.
According to this scenario, the burned gas expands and drives a flow
in the fuel mixture. The flow becomes nonuniform because of the
no-slip boundary condition at the wall which, together with
turbulence, distorts the flame front, increases the burning rate,
and leads to the acceleration. An accelerating flame front pushes
compression waves that continuously heat the fuel mixture ahead of
it until an explosion is triggered that eventually develops into a
detonation.

While the Shelkin mechanism relies prominently on the action of
turbulence, it was recently shown theoretically that flame
acceleration is possible even in its absence, in tubes with smooth
adiabatic wall \cite{Bychkov-et-al-2005,Akkerman-et-al-2006}. This
theory has been validated by extensive numerical simulations in
Refs. \cite{Bychkov-et-al-2005,Akkerman-et-al-2006} and supported by
experiments in smooth micro-tubes \cite{Wu.et.al-2007}. The theory
and modeling further demonstrate that laminar flame acceleration
becomes quite weak in wide tubes with increasing Reynolds number of
the flow, and as such loss to the wall may actually terminate the
process. Thus obstacles placed in the tube
\cite{Shelkin,Roy-et-al-2004,
Ciccarelli-Dorofeev-2008,Ciccarelli-et-al-2005,Kuznetsov-et-al-2005,
Frolov-et-al-2007,Johansen-Ciccarelli-2007,Gamezo-et-al-2007} appear
to be an essential factor in order to overcome the loss and
consequently support the DDT. It is generally believed that these
obstacles generate stronger turbulence, which increases the burning
rate and facilitates the flame acceleration.

However, in our recent work \cite{Bychkov-et-al-2008} we have
demonstrated that the obstacles play a more important role than just
producing turbulence, in that they provide a specific physical
mechanism of flame acceleration that is qualitatively different from
the Shelkin mechanism. This new mechanism is extra strong, providing
flame acceleration that is independent of the Reynolds number,
through the tube width, and as such may be quite important for
technical applications. Specifically, flame propagation in an
obstructed channel creates pockets of fresh fuel mixture between the
obstacles, as shown in Fig. \ref{fig-1}. Gas expansion due to
delayed burning in the pockets produces a powerful jet flow in the
unobstructed part of the channel. The jet flow renders the flame tip
to propagate much faster, which produces new pockets, generates a
positive feedback between the flame and the flow, and leads to flame
acceleration.  The accelerating flame reaches  supersonic speed with
respect to the tube wall, and consequently triggers explosion and
detonation.

The present paper extends the work of Bychkov et al., Ref.
\cite{Bychkov-et-al-2008}, in which the basic concepts of the new
mechanism were outlined, by presenting details of the theory and
simulations. Specifically, we discuss the influence of the
planar/axisymmetric flow geometry and gas compression on the
acceleration rate. We  show that the flame accelerates much faster
in cylindrical tubes than in planar channels, and that the
acceleration is slowed down due to gas compression. Furthermore, as
the Mach number of the flow increases, the acceleration process
saturates to statistically steady flame propagation at
supersonic speed with respect to the tube wall. This saturation state
may develop prior to the attainment of explosion and then
detonation. The flame speed in this state may be correlated with
the Chapman-Jouguet (CJ) deflagration speed
\cite{Landau&Lifshitz-1989,Chue-et-al-1993,Valiev-et-al-2009} and
with the state of fast flames observed experimentally
\cite{Ciccarelli-Dorofeev-2008,Kuznetsov-et-al-2002}. Finally, we
demonstrate numerically that flame acceleration may lead to
explosion and detonation triggering.

\section{2. Theory of flame acceleration}
\begin{figure}
\includegraphics[width=\columnwidth]{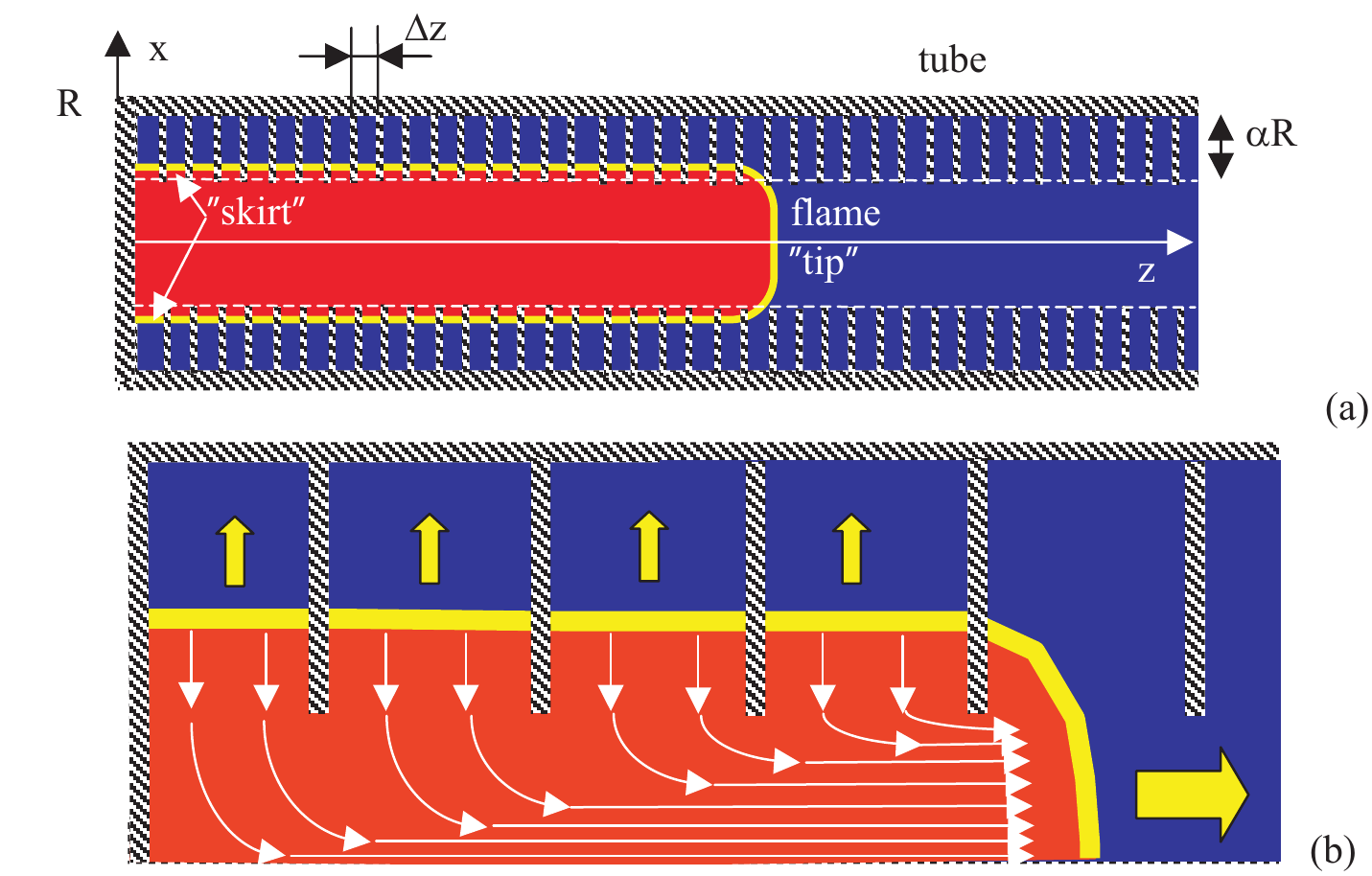}
\caption{Flame propagation in a channel with obstacles (a) and the
jet generation mechanism (b).}
 \label{fig-1}
\end{figure}

Figure \ref{fig-1} is a schematic of the problem under study. The
flame propagates from the closed end of a semi-infinite channel of
half-width (radius) $R$, with a fraction $\alpha < 1$ blocked by
obstacles. The central part of the channel, of half-width $(1 -
\alpha )R$, is unobstructed. The flame propagates extremely fast
along the unobstructed part of the channel, leaving behind pockets
of unburned mixture, between the obstacles, which will be burned
later. The deep narrow spaces between the obstacles, namely the
pockets, act as mini-channels in which the flame can be considered
to propagate mainly in the radial direction. This assumption is
most appropriate when the obstacles are placed close to each other
with deep pockets, $\Delta z << \alpha {\kern 1pt} R$, and with
slip boundary conditions at the wall. In general, burning in the
pockets depends on the obstacle geometry, whose influence will be
discussed later. We nevertheless recognize qualitatively the same
mechanism of flame acceleration in simulations and experiments
involving complicated obstacle shapes
\cite{Frolov-et-al-2007,Johansen-Ciccarelli-2007,Gamezo-et-al-2007}.

%============================= Figure 1 ============================%
%============================= Figure 1 ============================%

We now briefly state the fundamental concepts of the new mechanism
presented in Ref. \cite{Bychkov-et-al-2008}. We employ the standard
model of an infinitesimally thin flame front propagating normally to
itself with the unstretched laminar flame velocity $U_{f}$. The tube
wall is ideally slip and adiabatic. Non-slip boundary condition is
not needed for flame acceleration in the new mechanism and the
Reynolds number is not involved in the calculations. At the initial
stage of flame acceleration, the flow may be treated as
incompressible,
\begin{equation}
\label{eq1}
\nabla \cdot {\rm {\bf u}} = 0,
\end{equation}
while the fresh gas trapped in the pockets is burning. Expansion of
the burnt gas is characterized by the density ratio of the fuel
mixture and the burnt gas, $\Theta = \rho _{f} / \rho _{b} $, which
is quite large for most flames, with $\Theta = 5 - 8$. In the model
of a planar laminar flame front in the pockets, the fuel mixture in
a pocket is at rest, while the burnt gas is pushed out with the
velocity $(\Theta - 1)U_{f}$, see Fig. \ref{fig-1} (b). This value
determines the gas velocity at the border of the unobstructed part
of the channel
\begin{equation}
\label{eq1a} {\left| {u_{x}} \right|} = (\Theta - 1)U_{f} \quad
\textrm{at} \quad x = \pm (1 - \alpha )R.
\end{equation}
The shape of the flame tip is of minor importance for the present
mechanism, and it may be taken to be planar at all times.
Consequently, the flow of the burnt gas in the unobstructed channel
part is potential. Accounting for boundary condition (\ref{eq1a}), we find
the flow velocity
\begin{equation}
\label{eq2} (u_{x} ;\,u_{z} ) = {\frac{{(\Theta - 1)U_{f}}} {{(1 -
\alpha )R}}}(-x;\;z).
\end{equation}
The propagation speed of the infinitesimally thin flame front with
respect to the burnt gas is
\begin{equation}
\label{eq3}
{\frac{{dZ_{f}}} {{dt}}} - u_{z} (Z_{f} ) = \Theta U_{f} .
\end{equation}
Using the velocity distribution (\ref{eq2}), we find
\begin{equation}
\label{eq4} {\frac{{dZ_{f}}} {{dt}}} = {\frac{{(\Theta - 1)U_{f}}}
{{(1 - \alpha )R}}}Z_{f} + \Theta U_{f}.
\end{equation}
Integrating Eq. (\ref{eq4}) with the initial condition $Z_{f} (0) =
0$, we obtain a strong exponential acceleration of the flame tip
\begin{equation}
\label{eq5} {\frac{{Z_{f}}} {{(1 - \alpha )R}}} = {\frac{{\Theta}}
{{\Theta - 1}}}{\left[ {\exp (\sigma U_{f} t / R) - 1} \right]},
\end{equation}
with the scaled acceleration rate
\begin{equation}
\label{eq6}
\sigma = {\frac{{\Theta - 1}}{{1 - \alpha}} }.
\end{equation}
The derivation of Eqs. (\ref{eq2}) -- (\ref{eq6}) explains the basis
of the new acceleration mechanism presented in Ref.
\cite{Bychkov-et-al-2008}. This mechanism is quite powerful, with
the acceleration rate (\ref{eq6}) much larger than that of the
Shelkin mechanism in smooth tubes \cite{Bychkov-et-al-2005}. The
scaled acceleration rate does not depend on the Reynolds number, and
hence the viscosity and the tube width. As will be discussed below,
viscosity and turbulence may induce supplementary effects in the new
acceleration mechanism. As pointed out in Ref.
\cite{Bychkov-et-al-2008}, this new acceleration mechanism has many
features in common with the acceleration of finger flames, Refs.
\cite{Clanet-Searby-1996,Bychkov-et-al-2007}, although the finger
flame acceleration is quite limited in time, yielding maximum
increase in the burning rate by a factor of only 10-15 relative to
the planar flame speed \cite{Bychkov-et-al-2007}. In contrast, the
new physical mechanism leads to supersonic flame propagation with
respect to the tube wall, with possible transition to explosion and
detonation. It was also demonstrated in Ref.
\cite{Bychkov-et-al-2008} that the new mechanism remains effective
even when the flame skirt touches the main wall of the channel.
%============================= Figure 2 ============================%
\begin{figure}
\includegraphics[width=0.7\columnwidth]{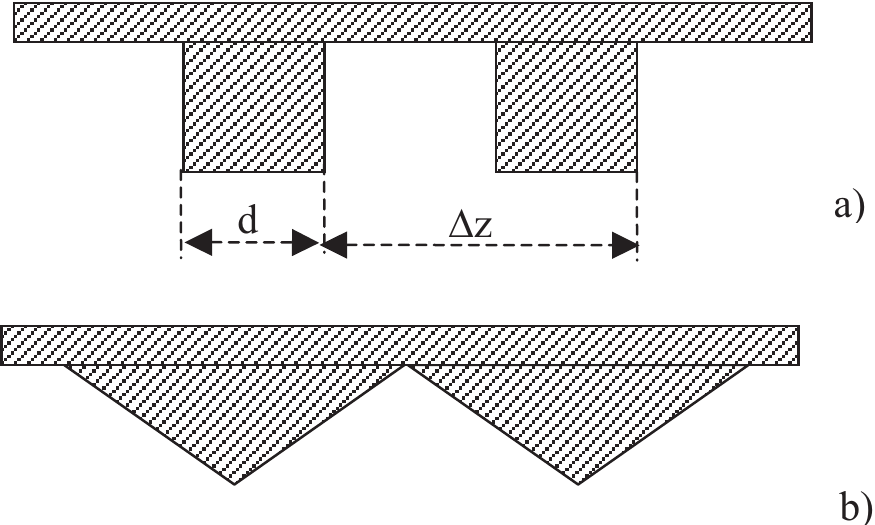}
\caption{Typical obstacle shapes discussed in ICDERS-2007.}
 \label{fig-4}
\end{figure}
%============================= Figure 2 ============================%

Theoretical understanding of the new acceleration mechanism enables
the analysis of the optimal obstacle geometry for flame
acceleration. Figure \ref{fig-4} shows typical obstacle shapes
discussed in Refs.
\cite{Frolov-et-al-2007,Johansen-Ciccarelli-2007,Johansen-Ciccarelli-2009,
Gamezo-et-al-2007,Gamezo-et-al-2008}. Specifically, in Fig.
\ref{fig-4} (a) we have rectangular obstacles of thickness $d$ and
spacing $\Delta z$. In this case delayed burning occurs only between
the obstacles, while the volume occupied by an obstacle itself is
ineffective for flame acceleration. Averaging volume production
in the burning process over the obstacle step, we rewrite the
boundary condition (\ref{eq1a}) as
\begin{equation}
\label{Eq.24-a}
\left| {{\left\langle {u_{x}}  \right\rangle}}  \right| = (1 - d / \Delta
z)(\Theta - 1)U_{f} \ \  \text{at} \ \  x = \pm (1 - \alpha )R,
\end{equation}
which leads to the reduced acceleration rate
\begin{equation}
\label{eq25}
\sigma = (1 - d / \Delta z){\frac{{\Theta - 1}}{{1 - \alpha}} }.
\end{equation}
Thus, thinner obstacles produce stronger acceleration. Another
possible factor suggested was obstacle phase shift at the bottom of
the channel in comparison to that at the top. This suggestion,
however, is not expected to substantially affect the laminar flame
dynamics, although it could influence the turbulence generated in
the flow in later stages. The triangular obstacle shape in Fig.
\ref{fig-4} (b) also renders the flame acceleration slower because
it reduces the volume of the fresh fuel mixture trapped between the
obstacles. Consequently, the optimal design for flame acceleration
is that of the infinitely thin obstacles shown in Fig. \ref{fig-1}.
For a fixed tube radius $R$, maximum acceleration rate is achieved
for maximum possible blockage ratio, i.e. for minimum value of
$(1-\alpha)R$. However, for a fixed width of the free channel part
$(1-\alpha)R$, the acceleration rate does not depend on the total
tube radius $R$, i.e. it does not depend on the depth of the
pockets. This is true, of course, only when pockets are sufficiently
deep with $\alpha$ comparable to unity. Very small obstacles
($\alpha \ll 1$) cannot be treated as obstacles and the new
mechanism is ineffective. Still, typical experimental configurations
employ obstacles with the blockage ratio of $1/4<\alpha <3/4$, for
which the present mechanism is quite effective. The dimensional
acceleration rate $(\Theta - 1)U_{f}/R(1-\alpha)$ is determined by
the expansion factor $\Theta$, the laminar flame velocity $U_{f}$
and the half-width of the unobstructed channel part $R(1-\alpha)$.
The theoretical model of a laminar flow does not predict any
dependence of the acceleration rate on the spacing between the
obstacles. As we shall demonstrate numerically in Sec. 4, the
spacing between the obstacles determines primarily the amplitude of
the velocity pulsations (see also Ref. \cite{Bychkov-et-al-2008}).
Still, these pulsations do not change the average acceleration rate,
which remains quite close to the predictions of the theoretical
model.

Additional increase in the acceleration rate is realized in the
axisymmetric geometry. Here we consider an axisymmetric tube with
obstacles in the form of planar rings blocking the space $(1 -
\alpha )R < r < R$ similar to Fig. \ref{fig-1}. Suppose the flame
tip accelerates as $Z_{f} = Z_{f} (t)$. A pocket between the
obstacles at the position $z$ starts burning at the instant $t_{f}
(z)$, where $t_{f} (z)$ is the inverted function $Z_{f} (t)$. The
flame in the axisymmetric pockets expands with the radius growing as
\begin{equation}
\label{eq26}
R_{f} = (1 - \alpha )R + U_{f} {\left[ {t - t_{f} (z)} \right]}.
\end{equation}
The radial velocity at the exit of the pocket, at $r = (1 - \alpha
)R$, is given for an incompressible flow as
\begin{equation}
\label{eq27}
(\Theta - 1)R_{f} U_{f} = - (1 - \alpha )Ru_{r} ,
\end{equation}
which determines the boundary condition at the border of the unobstructed
part of the channel $r = (1 - \alpha )R$,
\begin{equation}
\label{eq28}
u_{r} = - (\Theta - 1)U_{f} \left( {1 + {\frac{{U_{f}}} {{(1 - \alpha
)R}}}{\left[ {t - t_{f} (z)} \right]}} \right).
\end{equation}
In the event of exponential, or near-exponential, flame acceleration
\begin{equation}
\label{eq29} Z_{f} = Z_{0} {\left[ {\exp (\sigma U_{f} t / R) - 1}
\right]},
\end{equation}
the instant $t_{f} (z)$ varies logarithmically, and hence slowly,
with z, as
\begin{equation}
\label{eq30}
t_{f} = {\frac{{R}}{{\sigma U_{f}}} }\ln (z / Z_{0} + 1),
\end{equation}
where $Z_{0} $ is some amplitude.

Let us first determine the flame acceleration neglecting the
increase in the flame radius $R_{f} $ of Eq. (\ref{eq26}) in
comparison with $(1 - \alpha )R$. Then, the axisymmetric solution to
the continuity equation (\ref{eq1}) in the unobstructed part of the
channel is
\begin{equation}
\label{eq31}
(u_{r} ;\,u_{z} ) = {\frac{{(\Theta - 1)U_{f}}} {{(1 - \alpha )R}}}( -
r;\;2z),
\end{equation}
which is the axisymmetric counterpart of Eq. (\ref{eq2}). The
solution (\ref{eq3}) -- (\ref{eq6}) is respectively modified in the
axisymmetric geometry as
\begin{equation}
\label{eq32} {\frac{{dZ_{f}}} {{dt}}} = 2{\frac{{(\Theta -
1)U_{f}}}{{(1 - \alpha )R}}}Z_{f} + \Theta U_{f},
\end{equation}
with
\begin{equation}
\label{eq33} {\frac{{Z_{f}}} {{(1 - \alpha )R}}} = {\frac{{\Theta}}
{{2(\Theta - 1)}}}{\left[ {\exp (\sigma U_{f} t / R) - 1} \right]},
\end{equation}
and
\begin{equation}
\label{eq34}
\sigma = 2{\frac{{\Theta - 1}}{{1 - \alpha}} }.
\end{equation}
Equation (\ref{eq34}) shows that flame acceleration in the
axisymmetric geometry is twice that in the planar case.

We next account for the increase in $R_{f}$ as compared to $(1 -
\alpha)R$. Taking the radial velocity component in the same form as
in Eq. (\ref{eq31}),
\begin{equation}
\label{eq35} u_{r} = - {\frac{{(\Theta - 1)U_{f}}} {{(1 - \alpha
)R}}}r\left({1 + {\frac{{U_{f}}} {{(1 - \alpha )R}}}{\left[{t -
t_{f} (z)} \right]}} \right),
\end{equation}
we find the respective z-velocity component
\begin{equation}
\label{eq36}
u_{z} = 2{\frac{{(\Theta - 1)U_{f}}} {{(1 - \alpha )R}}}\left( {z +
{\frac{{U_{f}}} {{(1 - \alpha )R}}}\int {{\left[ {t - t_{f} (z)}
\right]}\,}
dz} \right),
\end{equation}
and its value at the flame tip
\begin{equation}
\label{eq37} u_{z} = 2{\frac{{(\Theta - 1)U_{f}}} {{(1 - \alpha
)R}}}Z_{f} \left( {1 + {\frac{{U_{f}}} {{(1 - \alpha
)R}}}{\left\langle {t - t_{f} (z)} \right\rangle}} \right),
\end{equation}
where
\begin{equation}
\label{eq38}
{\left\langle {t - t_{f} (z)} \right\rangle}  = t - {\frac{{1}}{{Z_{f}
}}}{\int\limits_{0}^{z_{f}}  {t_{f} (z)\,dz}} .
\end{equation}

When the fuel mixture in the first pocket is almost completely
burnt, accounting for Eqs. (\ref{eq30}), (\ref{eq33}), the averaging
in Eq. (\ref{eq38}) yileds
\begin{equation} \label{eq39} \sigma _{0}
{\frac{{U_{f}}} {{R}}}{\left\langle {t - t_{f} (z)} \right\rangle}
= 1 - {\frac{{Z_{0}}} {{Z_{f}}} }\ln \left( {{\frac{{Z_{f}
}}{{Z_{0}}} } + 1} \right),
\end{equation}
where $\sigma _{0} $ is given by Eq. (\ref{eq34}). The second term
in Eq. (\ref{eq39}) diminishes asymptotically to zero with time, but
the first term leads to corrections to Eqs. (\ref{eq32}),
(\ref{eq34}) as
\begin{equation}
\label{eq40} {\frac{{dZ_{f}}} {{dt}}} = 2{\frac{{(\Theta - 1)U_{f}}}
{{(1 - \alpha)R}}}\left( {1 + {\frac{{1}}{{2(\Theta - 1)}}}}
\right)Z_{f} + \Theta U_{f},
\end{equation}
and
\begin{equation}
\label{eq41}
\sigma = 2{\frac{{\Theta - 1}}{{1 - \alpha}} }\left( {1 +
{\frac{{1}}{{2(\Theta - 1)}}}} \right).
\end{equation}
The second term in Eq.~(\ref{eq41}) may be treated as corrections to
the first one in the limit of $2(\Theta - 1) > > 1$, which holds
with good accuracy of about 7\% for realistic fuel mixtures.
Furthermore, this correction becomes important only when the leading
part of the flame skirt has almost touched the main wall, when
burning in the first pocket is almost finished. At the beginning of
the burning process, the acceleration rate should be approximated by
Eq.~(\ref{eq34}). However, even during this initial period of flame
acceleration, the flame velocity may attain fairly large values
relative to the sound speed, at which the incompressibility
assumption fails. Thus, the flame acceleration rate should be
evaluated by Eq.~(\ref{eq34}) rather than Eq.~(\ref{eq41}).
%============================= Figure 3============================%
\begin{figure}
\includegraphics[width=\columnwidth]{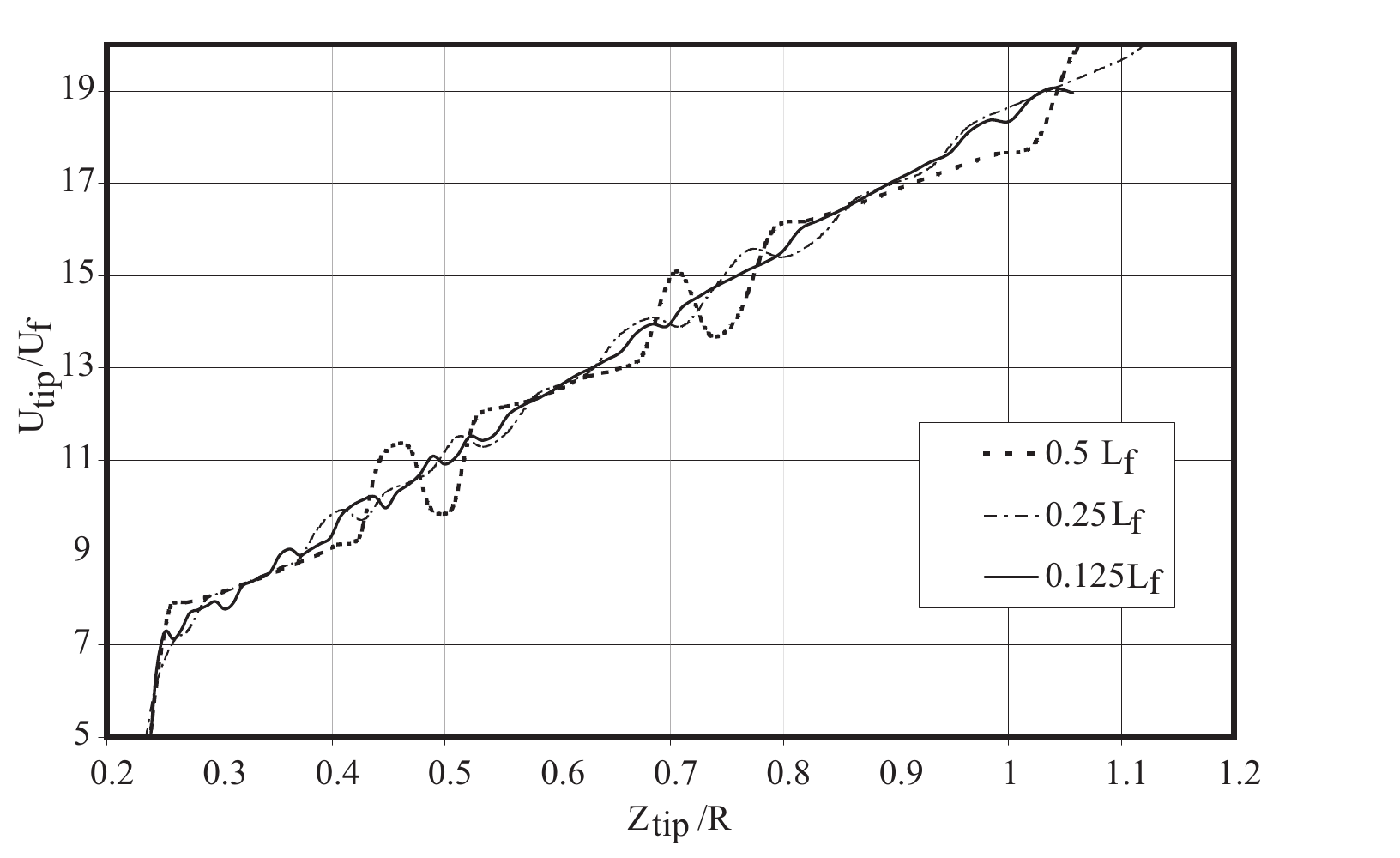}
\caption{Resolution test: flame tip velocity versus tip position for
$Ma=0.001$, $\alpha=1/2$ and various mesh sizes
$(0.125;0.25;0.5)L_{f}$.}
 \label{fig-5}
\end{figure}
%============================= Figure 3============================%

The theory above constitutes the backbone of the new mechanism of
extremely fast flame acceleration in tubes/channels with obstacles.
Still, this theory may be developed further to incorporate other
effects, such as gas compression, viscosity, non-slip at the wall,
etc, which also influence the acceleration process. Here, we discuss
briefly some of these effects.
%============================= Figure 4 ============================%
\begin{figure}
\includegraphics[width=\columnwidth]{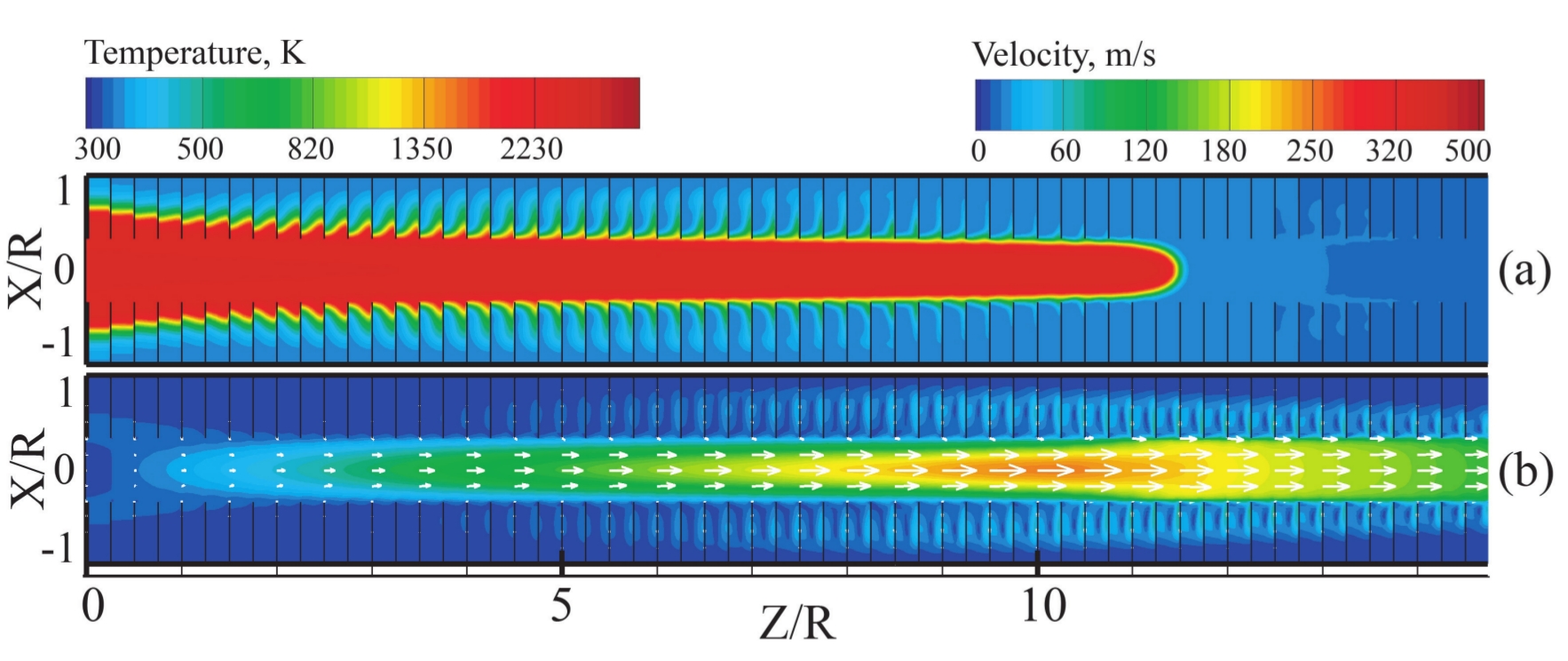}
\includegraphics[width=\columnwidth]{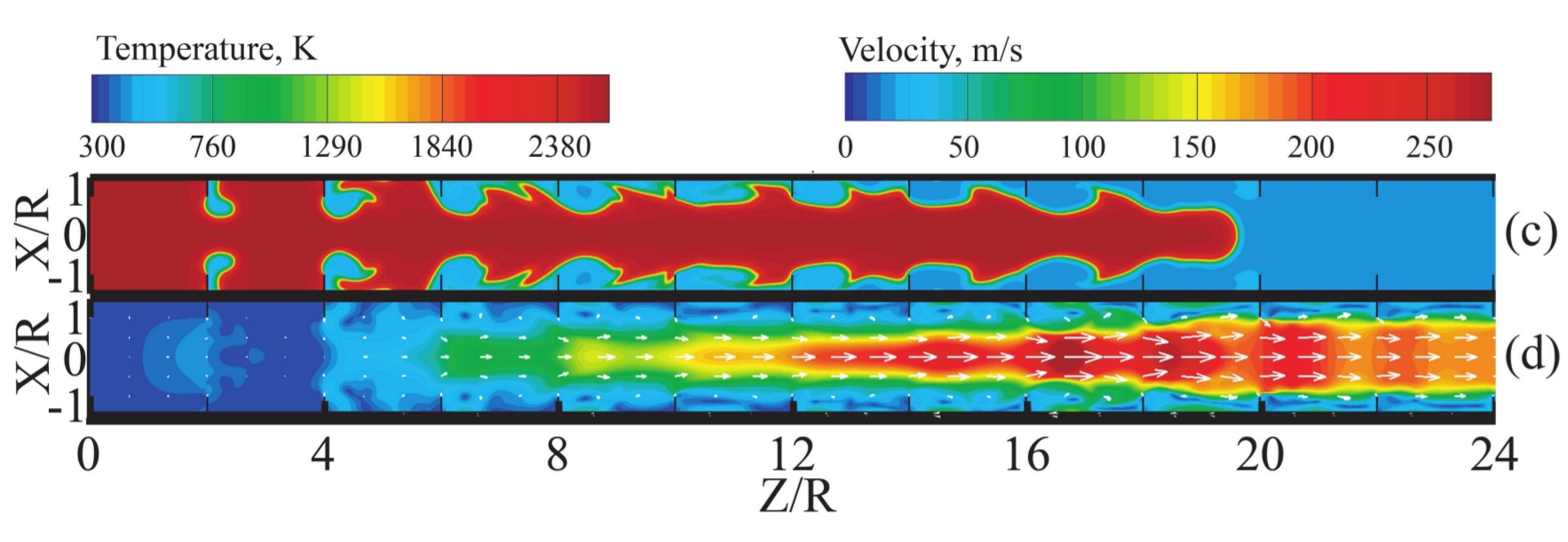}
\includegraphics[width=\columnwidth]{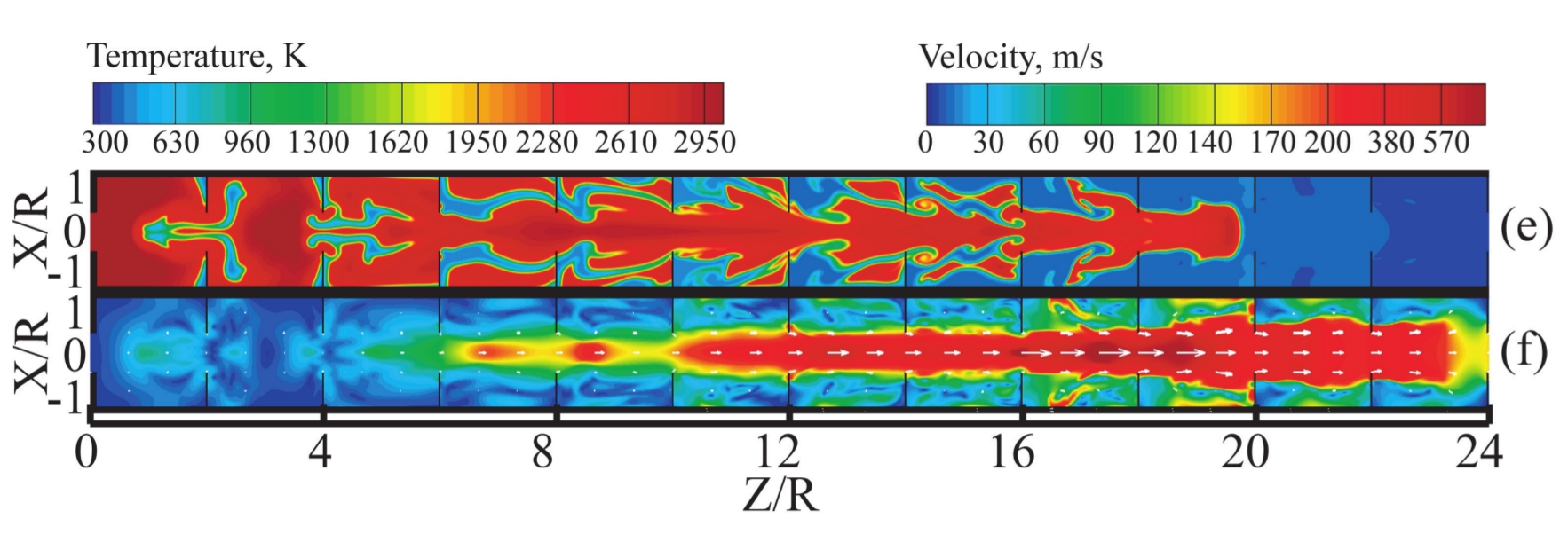}
\includegraphics[width=\columnwidth]{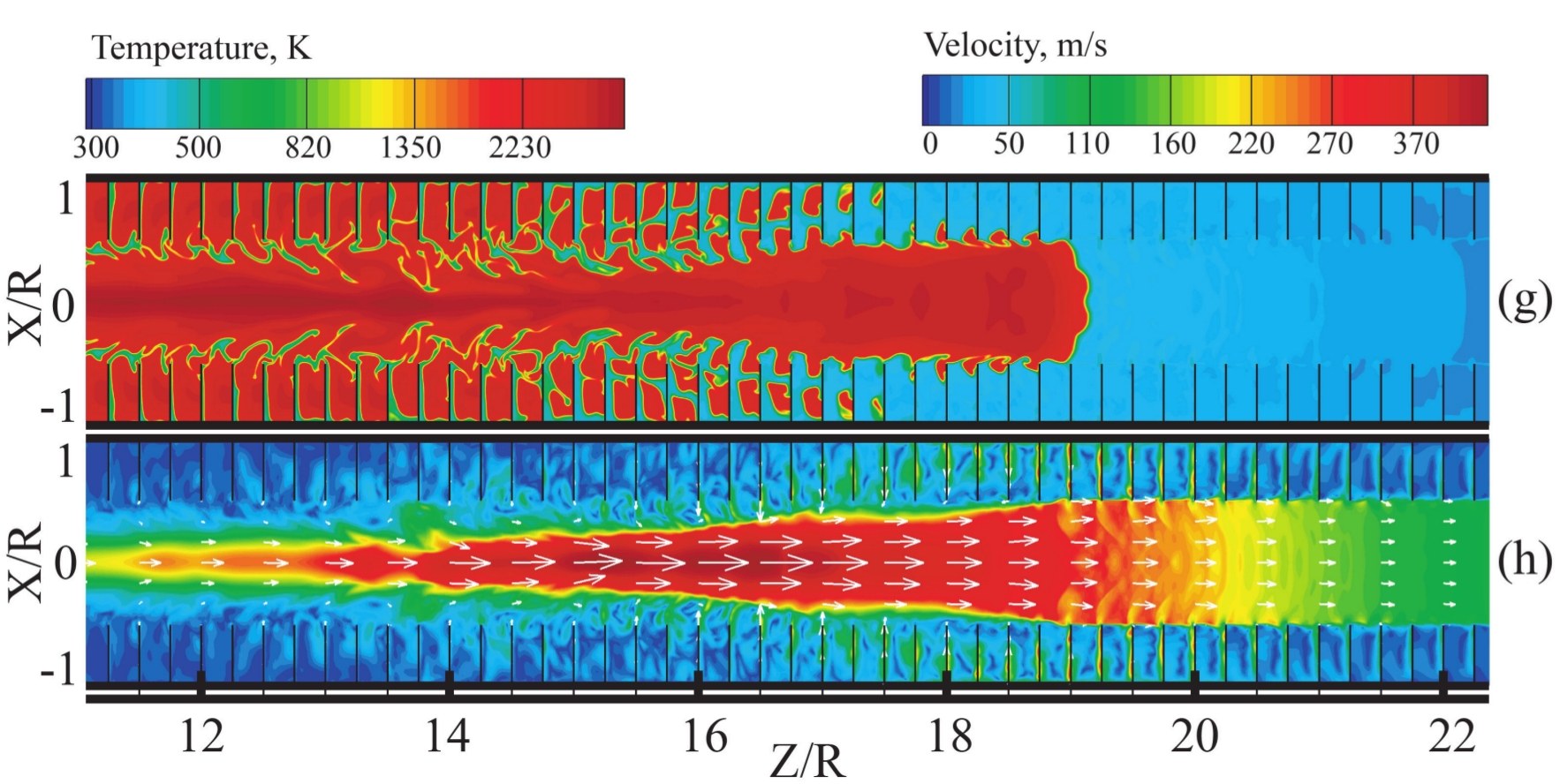}
\caption{ Snapshots of temperature and velocity in the flow
generated by an accelerating flame in the planar geometry for the
following parameters: (a, b) $Ma = 0.001$, $\Delta z / R = 1 / 4$,
$\alpha = 2 / 3$ (appears in \cite{Bychkov-et-al-2008}); (c, d) $Ma
= 0.005$, $\Delta z / R = 2$, $\alpha = 1 / 3$; (e, f) $Ma = 0.005$,
$\Delta z / R = 2$, $\alpha = 2 / 3$; (g, h) $Ma = 0.001$, $\Delta z
/ R = 1 / 4$, $\alpha = 1 / 2$ (appears in
\cite{Bychkov-et-al-2008}).}
 \label{fig-6}
\end{figure}
%============================= Figure 4 ============================%

(1) Gas compression: The present theory is developed
for incompressible flows. Compressibility is expected to moderate
the flame acceleration considerably by recognizing that the flame
velocity is limited by the CJ deflagration value. Preliminary
numerical results on this subject are given in Ref.
\cite{Bychkov-et-al-2008}, and similar effects for smooth tubes
are also discussed in Ref. \cite{Valiev-et-al-2009}. We shall
perform additional modeling later to demonstrate the role of gas
compression.

(2) Curved flame shape in the pockets and at non-slip wall:
Flame shape in the pockets is not necessarily planar, as we assumed
in the calculations. Inclined or curved flame shapes provide faster
propagation of the front and stronger cumulative expansion. A flame
may acquire a curved shape because of intrinsic instabilities,
viscous effects, etc. For example, the viscosity and non-slip wall
can increase the flame speed by a factor of about~1.5
\cite{Akkerman-et-al-2006a}. The curved flame shape in the pockets
renders acceleration of the flame tip stronger. In such cases the
planar flame velocity in Eq. (\ref{eq1a}) should be multiplied by some factor
describing the increase in the flame velocity in the pockets.

(3) Turbulence in the pockets: Flow in the unobstructed part
of the channel pushed by the accelerating flame also generates
turbulence in the pockets, see, for example,
\cite{Bychkov-et-al-2008,Ciccarelli-Dorofeev-2008,
Johansen-Ciccarelli-2007}. Turbulence corrugates the flame front and
increases the burning rate in the pockets. The effect of turbulence
becomes especially noticeable when the distance between the
obstacles is comparable to or larger than the obstacle size, $\Delta
z \propto \alpha {\kern 1pt} R$.

(4) Large spacing  between the obstacles: Large spacing
$\Delta z$ causes two competing effects. On the one hand, the
flow in the unobstructed part of the channel becomes less confined,
which may be interpreted as decreased effective blockage ratio and
thereby reduces the acceleration rate. On the other hand, large
spacing leads to stronger turbulence in the pockets thus augmenting
flame acceleration. Numerical simulation \cite{Bychkov-et-al-2008}
showed that these effects compensate each other at moderate values of
$\Delta z/R=1, \ 2$, with turbulence having a slightly stronger effect
in the developed stage of flame acceleration.

(5) Loss to the wall: In smooth tubes heat loss to the wall
reduces expansion of the burning gas dramatically and slows down
flame acceleration. However, in tubes with sufficiently deep and
thin obstacles this negative influence may be reduced considerably.
Indeed, thin obstacles may be heated rapidly and do not extract
much energy from the flow. At the same time, loss to the main wall
is minimized since it is mostly in contact with the fresh fuel
mixture instead of the burnt gas. One should expect that loss to the
main wall is equivalent to an effective smaller radius of the tube,
$R_{eff} < R$, but with the same size of the unobstructed part of
the channel $(1 - \alpha )R$. Since the acceleration rate is
determined by $(1 - \alpha )R$, not by $R$, then one should expect
minimal influence of the loss in such a configuration.

\section{3. Basic equations and numerical method}
We performed direct numerical simulation of the hydrodynamic and
combustion equations including transport processes and Arrhenius
kinetics. Both 2D planar and axisymmetric cylindrical flows were
investigated using the Navier-Stokes system of the governing
equations presented, e.g., in Refs.
\cite{Bychkov-et-al-2005,Akkerman-et-al-2006}. To avoid the
thermal-diffusive instability we took unity Lewis number $Le  = 1$,
with $\Pr  = 0.75$ and the dynamical viscosity  $\mu = 1.7\times
10^{ - 5}{\rm N}{\rm s}{\rm /} {\rm m}^{{\rm 2}}$. The fuel--air
mixture and burnt gas were assumed to be a perfect gas with a
constant molar mass $m = 2.9\times 10^{ - 2}{\rm k}{\rm g}{\rm /}
{\rm m}{\rm o}{\rm l}$. We considered a one-step irreversible
Arrhenius reaction with an activation energy $E_{a}$,
pre-exponential factor of inverse time dimension $\tau _{R}^{-1}$,
first order dependence on concentration of the fuel mixture, and
first or second order dependence on density similar to Ref.
\cite{Kagan-et-al-2006}. The first order was used in all studies of
flame acceleration and saturation to the CJ state. The second order
was employed to obtain explosion and transition to detonation, since
in that case flame dynamics is much more sensitive to pressure and
temperature build-up in the compression waves generated by an
accelerating flame. In the simulations we took $E / R_{p} T_{f} =
32$ in order to have better resolution of the reaction zone. In the
present study we focus mainly on the flame acceleration and
preheating of the fuel mixture ahead of the flame front, which do
not depend on the activation energy. The activation energy is
crucial for time and position of explosion triggering and DDT.
However, proper \textsl{quantitative} investigation of DDT cannot be
performed within a simplified one-step mechanism of chemical
reaction kinetics. Instead, it requires separate description of the
chemical kinetics at high temperatures (flame, detonation) and low
temperatures (explosion ignition). For this reason, in the present
work we perform only qualitative study of explosion triggering
choosing a scaled activation energy convenient for such a study. The
factor $\tau_{R}$ was adjusted to obtain a particular value of the
planar flame velocity $U_{f}$ by solving the associated eigenvalue
problem \cite{Travnikov_1997,Modestov-et-al-09}. For example, taking
the planar flame velocity $U_{f} = 34.7\ \textrm{cm/s}$, we  set
$\tau _{R} = 4.06\cdot 10^{-8}\ \textrm{s}$ for the first-order
reaction. The flame thickness is conventionally defined as
\begin{equation}
\label{eq56}
L_{f} \equiv {\frac{{\mu _{f}}} {{\Pr \rho _{f} U_{f}}} },
\end{equation}
where $\rho _{f} = 1.16\,kg / m^{3}$ is the unburnt mixture density.
We took initial temperature of the fuel mixture $T_{f} = 300K$,
initial pressure $P_{f} = 10^{5}Pa$, adiabatic index $\gamma = 1.4$,
and initial expansion ratio $\Theta = 8$. We took different values
of the initial Mach number in the range $10^{ - 3} \le Ma \equiv
U_{f} / c_{s} \le 10^{ - 2}$, with the lower limit corresponding to
realistic methane and propane flames. By varying the Mach number, we
investigated the influence of gas compression on flame acceleration.
The theory of Sec. 2 does not involve the Reynolds number, which
implies minor dependence of the results on the channel/tube width,
provided it is sufficiently large. On the other hand, simulation of
burning in wide channels with obstacles can be quite time consuming.
For this reason, in the simulations we used a moderate value of the
channel half-width (radius), $R = 24L_{f}$. Some of the 2D planar
simulation runs were performed only for half of the channel assuming
symmetry. However, the majority of runs were for the entire channel,
which is especially important for the developed stages of flame
acceleration involving turbulence as well as for explosion triggering.

%============================= Figure 5 u-axial ============================%
\begin{figure}
\includegraphics[width=\columnwidth]{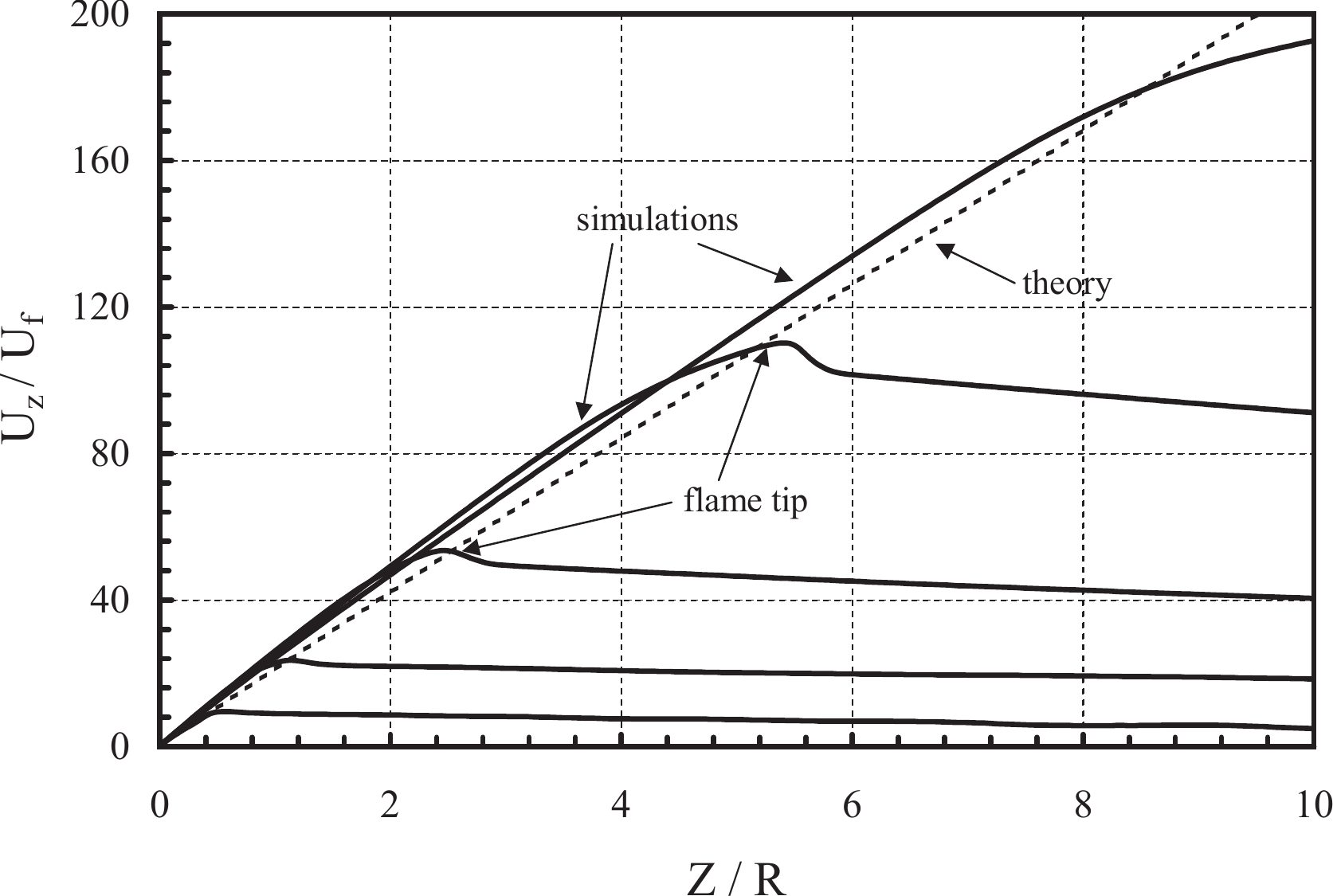}
 \caption{ Profile of the scaled z-velocity along the channel axis plotted
 for $\alpha = 2/3$, $Ma = 0.001$, $\Delta z / R = 1 / 4$ at the time
 instants $U_{f}t/R=0.02-0.16$ equally spaced in time. The theoretical line
 is related to Eq. (\ref{eq3}).}
\label{fig-u-axial}
\end{figure}
%============================= Figure 5 u-axial ============================%
%============================= Figure 6 u-x ================================%
\begin{figure}
\includegraphics[width=\columnwidth]{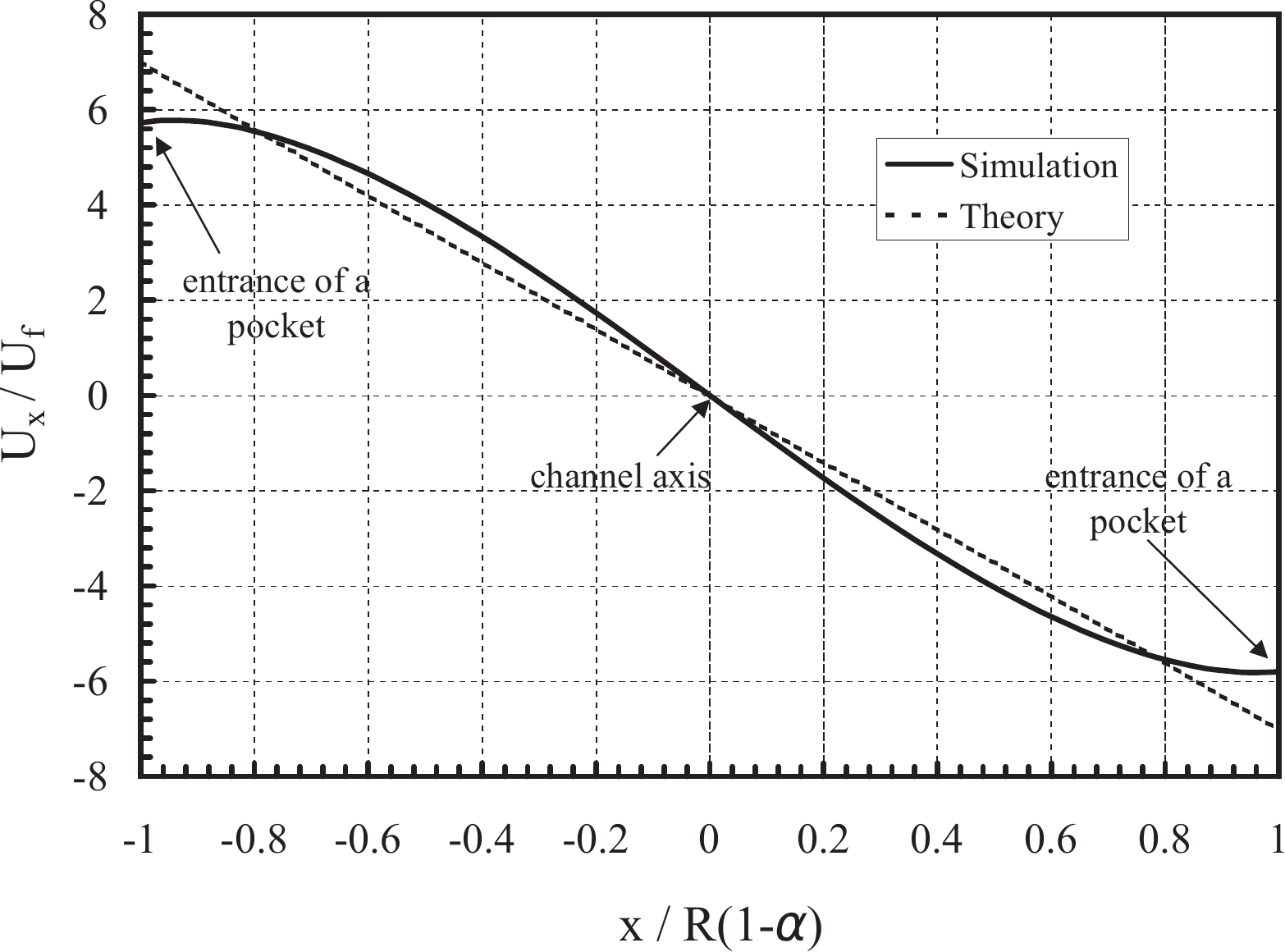}
 \caption{ Profile of the scaled x-velocity along the cross-section $z/R=0.5$
 plotted for $\alpha = 2/3$, $Ma = 0.001$, $\Delta z / R = 1 / 4$ at the
 time instant $U_{f}t/R=0.16$. The theoretical line is related to Eq.
 (\ref{eq3}).}
\label{fig-u-x}
\end{figure}
%============================= Figure 6 u-x ================================%

The present modeling is also relevant to combustion in
micro-channels, which is a rapidly developing subject. Recent
theory, modeling and experiments
\cite{Bychkov-et-al-2005,Akkerman-et-al-2006,Wu.et.al-2007}
demonstrated the possibility of DDT in smooth micro-tubes. The
present results also successfully predict flame acceleration and DDT
in micro-channels with obstacles, while recognizing that the present
analysis of course has a wider implementation than micro-channel
flows. According to the theory of Sec. 2, the same extra-strong
mechanism of flame acceleration works even in very wide tubes,
though it is impossible to attain such a state in numerical
simulations at present because of the inevitable computational
limitations. The channel width employed in the present simulation
determines the Reynolds number related to the laminar flame speed
$Re = U_{f} R / \nu = R / L_{f} \Pr = 32$. The Reynolds number
related to the flow $Re = {\left\langle {u_{z}} \right\rangle} 2R /
\nu $ can be larger by several orders of magnitude due to flame
acceleration and thermal expansion of the burnt gas. Flame
acceleration and increase in the Reynolds number may produce
turbulence in the flow. Indeed, our simulation shows that the
burning happens in the laminar regime at the beginning of flame
acceleration, and considerable turbulence is generated close to the
end of the process. Turbulence level depends typically on the
obstacle spacing.

%============================= Figure 7 ====================================%
\begin{figure}
\includegraphics[width=\columnwidth\centering]{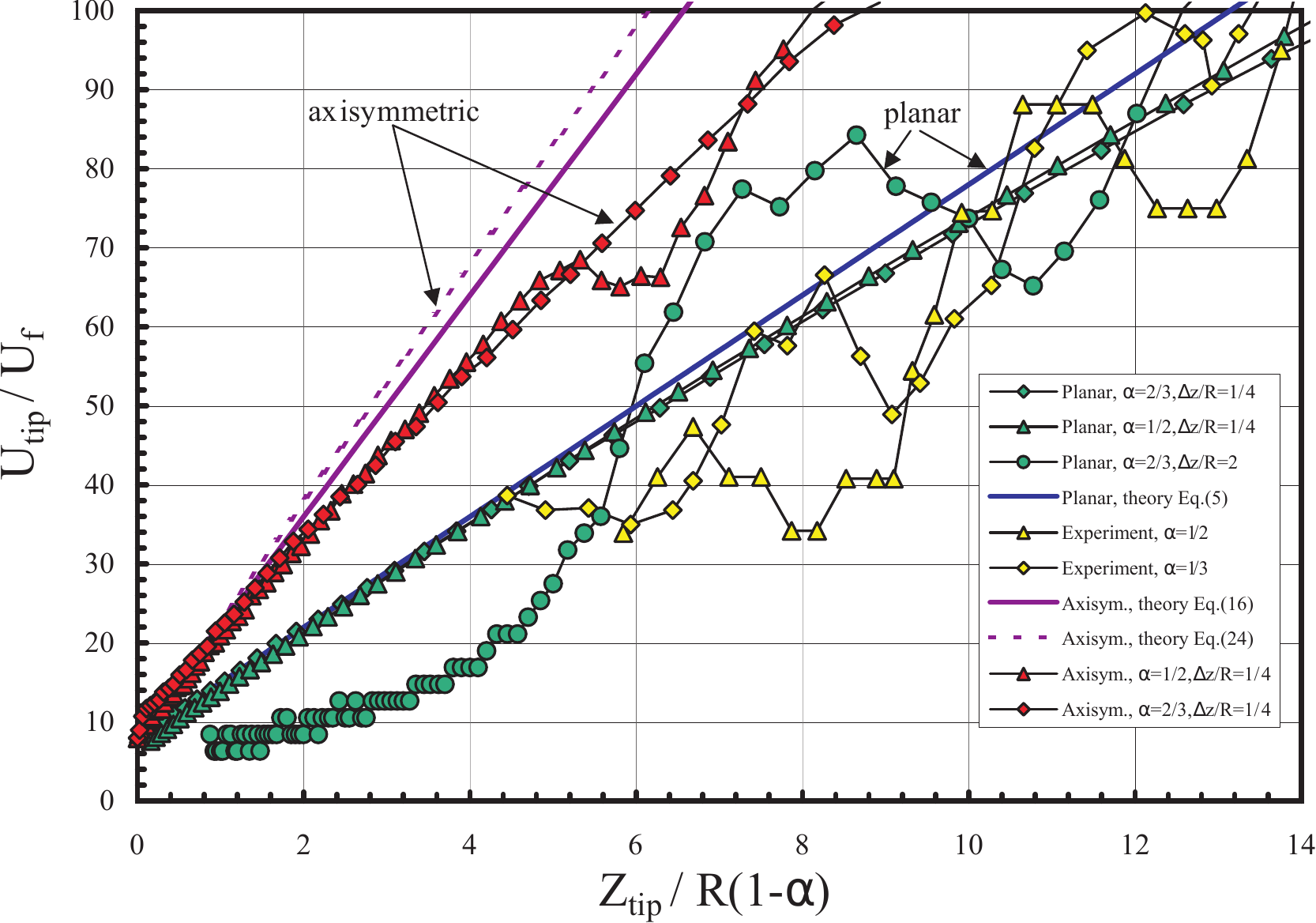}
 \caption{ Scaled velocity of the flame tip versus the scaled tip position
 for planar and axisymmetric geometry. The theoretical lines correspond to
 Eqs. (\ref{eq4}), (\ref{eq32}), (\ref{eq40}). The symbols show results of
 numerical modeling for  $\alpha = 1/2;\, 2/3$ and $\Delta z / R = 1 / 4; \,2$.
 The experimental data shows results of Ref.~\cite{Johansen-Ciccarelli-2009}.}
 \label{fig-7}
\end{figure}
%============================= Figure 7 ====================================%

We took slip and adiabatic boundary conditions at the tube wall:
\begin{equation}
\label{eq57}
{\rm {\bf n}} \cdot {\rm {\bf u}} = 0,
\quad
{\rm {\bf n}} \cdot \nabla T = 0,
\end{equation}
where ${\rm {\bf n}}$ is the unit normal vector at the wall. At the
open end of the channel, non-reflecting boundary conditions were
used. As initial conditions, we used a hemisphere of hot "burnt" gas
at the channel axis at the closed end of the tube, with the
temperature profile given by the analytical solution of Zel'dovich
and Frank-Kamenetskii~\cite{Zeldovich.et.al-1985,Law},
\begin{eqnarray}
T & = & T_{f} + (T_{b} - T_{f} )\exp \left( { - \frac{\sqrt {x^{2} + z^{2}}}{ L_{f}}}
\right) \textrm{if} \   z^{2} + x^{2} < r_{f}^{2},  \nonumber \\
T & = & \Theta T_{f} \ \ \textrm{if} \ \ z^{2} + x^{2} > r_{f}^{2}, \nonumber \\
Y & = & (T_{b} - T) / (T_{b} - T_{f} ),  \ P = P_{f}, \ u_{x} = 0, \ u_{z} = 0. 
\end{eqnarray}
Here $r_{f}$ is the radius of the hemisphere. The boundary of the
hot gas is not a flame front yet, and we take both cold and hot gas
initially at rest. The finite initial radius $r_{f}$ is equivalent
to a time shift, which requires proper adjustments when comparing
the theory and numerical simulations. When necessary, we shifted the
numerical solution in time to have the theory and the modeling
results starting at the same point.

%============================= Figure 8 ====================================%
\begin{figure}[b]
\includegraphics[width=\columnwidth]{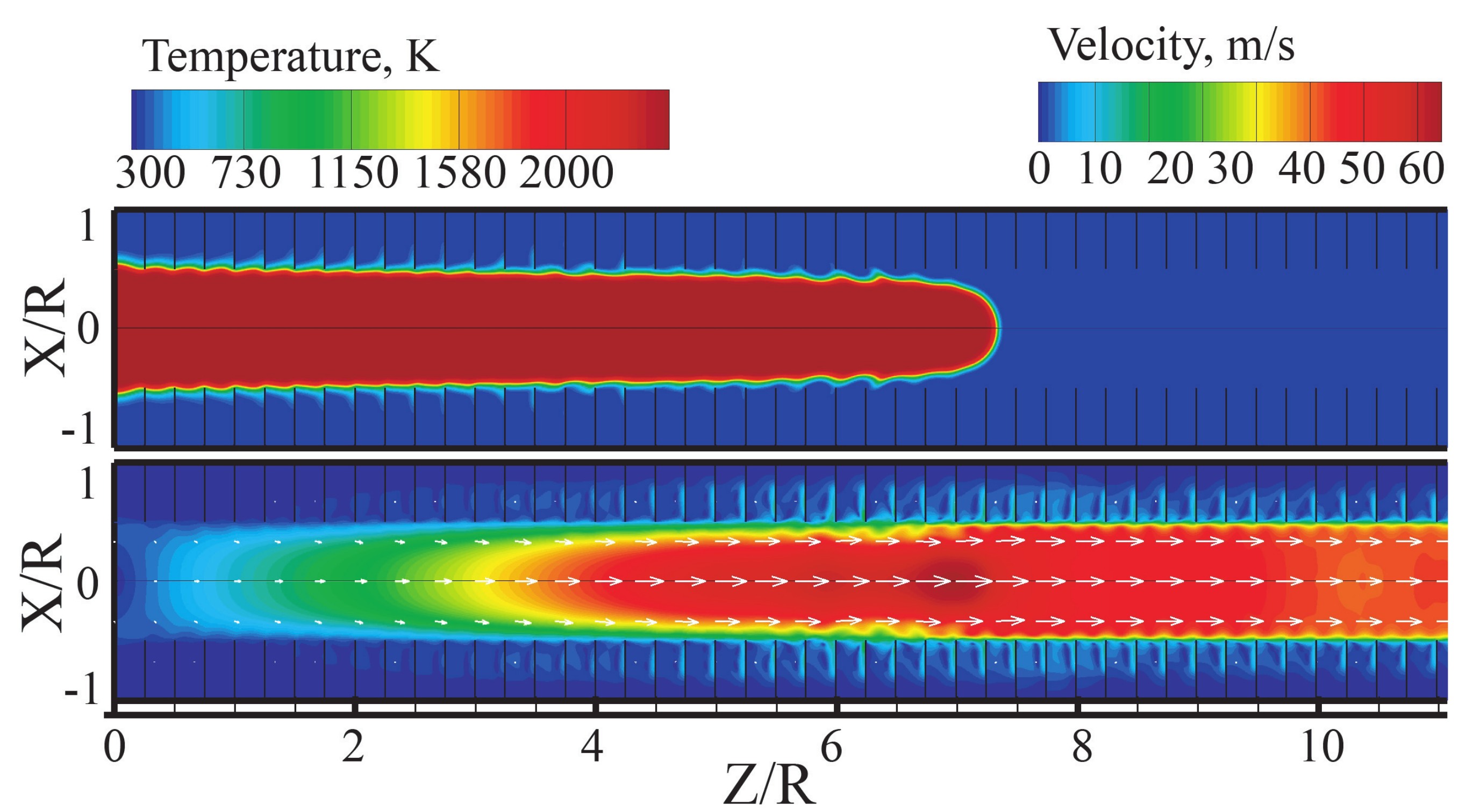}
\caption{ Snapshots of temperature and velocity in the flow
generated by an accelerating flame in an axisymmetric geometry for
$Ma = 0.001$, $\Delta z / R = 1 / 4$; $\alpha = 1 / 2$.}
 \label{fig-8}
\end{figure}
%============================= Figure 8 ====================================%

A 2D hydrodynamic Navier-Stokes code adapted for parallel
computation \cite{Eriksson-2005a,Eriksson-2005b,Eriksson-2006} was
used. The numerical scheme is second-order accurate in time, fourth
order in space for the convective terms, and second order in space
for the diffusive terms. The code is robust and accurate, having
been successfully used in combustion and aero-acoustic applications.
The code is available in 2D (Cartesian and cylindrical axisymmetric)
and 3D Cartesian versions. In the present work, we performed only 2D
simulations to save computational time and to be able to perform a
large number of simulation runs required for a thorough
investigation of the problem.

%============================= Figure 9 ====================================%
\begin{figure}
\includegraphics[width=\columnwidth]{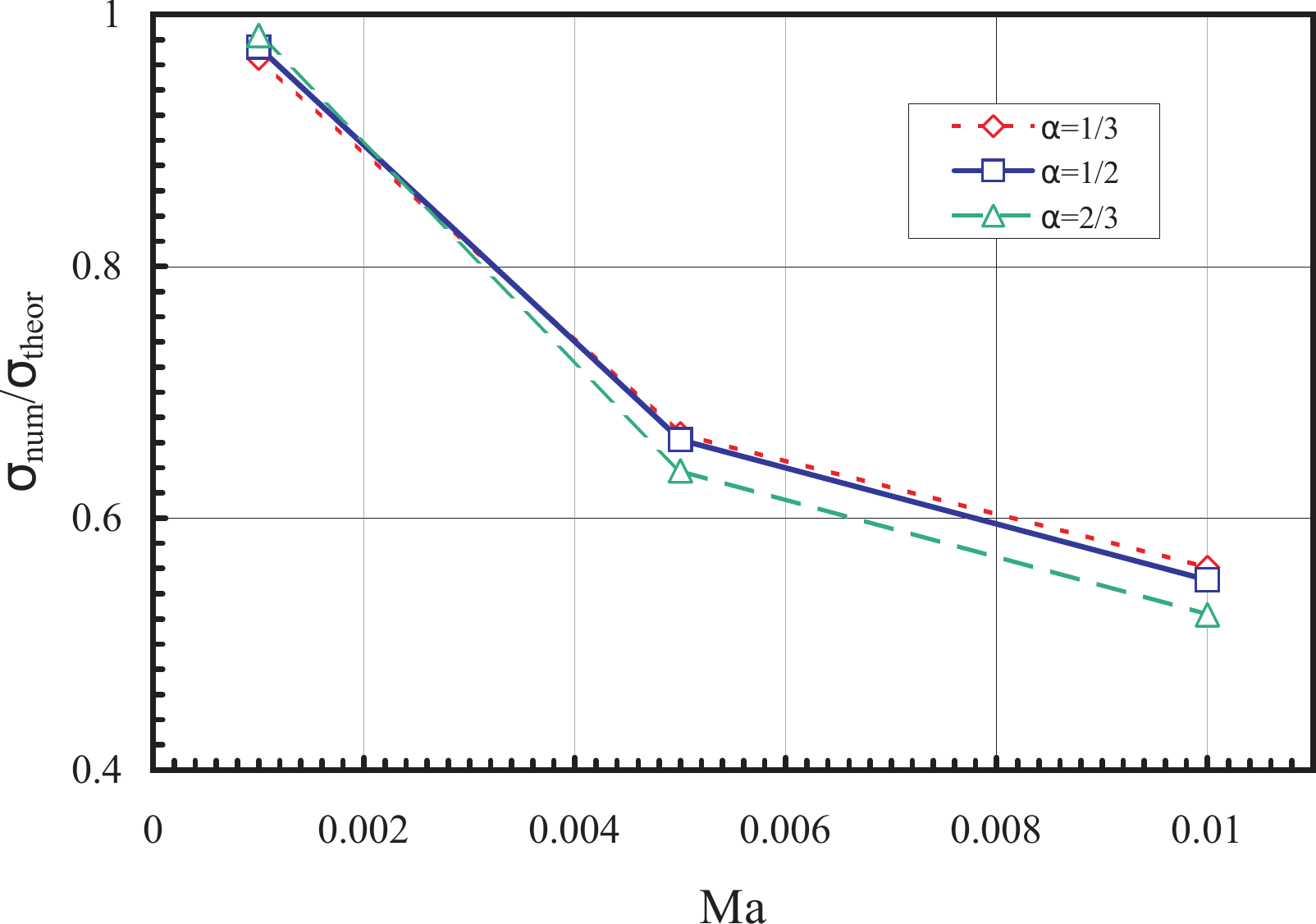}
\caption{ Ratio of the acceleration rate obtained numerically to
predicted theoretically, Eq. (7), versus the initial Mach number for
the planar geometry and  $\alpha = 1 / 3;\,1 / 2;\,2 / 3$ and
$\Delta z / R = 1 / 4$.}
 \label{fig-9}
\end{figure}
%============================= Figure 9 ====================================%

A uniform grid with quadratic cells of size $0.2L_{f} $ was used to
ensure isotropic propagation of the curved flame in x and y
directions. The longitudinal size of the calculation domain changes
dynamically, following the leading pressure wave. Splines of the
third order were used for re-interpolation of the flow variables
during periodic grid reconstruction to preserve second-order
accuracy of the numerical scheme. We performed several test
simulation runs with resolutions of 0.125$L_{f}$, 0.25$L_{f}$,
0.5$L_{f}$. The test demonstrated that the flame velocity grows
exponentially for all chosen resolutions, with the difference in the
scaled acceleration rate $\sigma $ not exceeding 6\%. The dependence
of the scaled flame tip velocity $U_{tip} / U_{f} $ on the scaled
distance $Z_{tip} / R$ is shown in Fig. \ref{fig-5} for different
resolutions. Resolution tests also showed convergence of the
acceleration rate $\sigma $ value with increasing resolution.

%Special care has been taken for the simulation runs, which included
%explosion and DDT. Temperature distribution is strongly nonuniform
%in a turbulent flow of the fuel mixture ahead of the flame front. On
%the other hand, explosion triggering is temperature sensitive and
%starts typically in the so-called "hot points" provided that these
%points are sufficiently large. In order to observe explosion and DDT
%we included addition numerical diffusion which provided proper
%spreading of temperature. At the same time, we checked that this
%numerical diffusion does not change properties either of the planar
%laminar flame front or of the flame acceleration, which is the main
%subject of the present paper. A similar addition numerical diffusion
%has been used, for example, in Ref. \cite{Gamezo-et-al-2008}.

\section{4. Simulation results and discussion}
%============================= Figure 10 ===================================%
\begin{figure}
\includegraphics[width=\columnwidth]{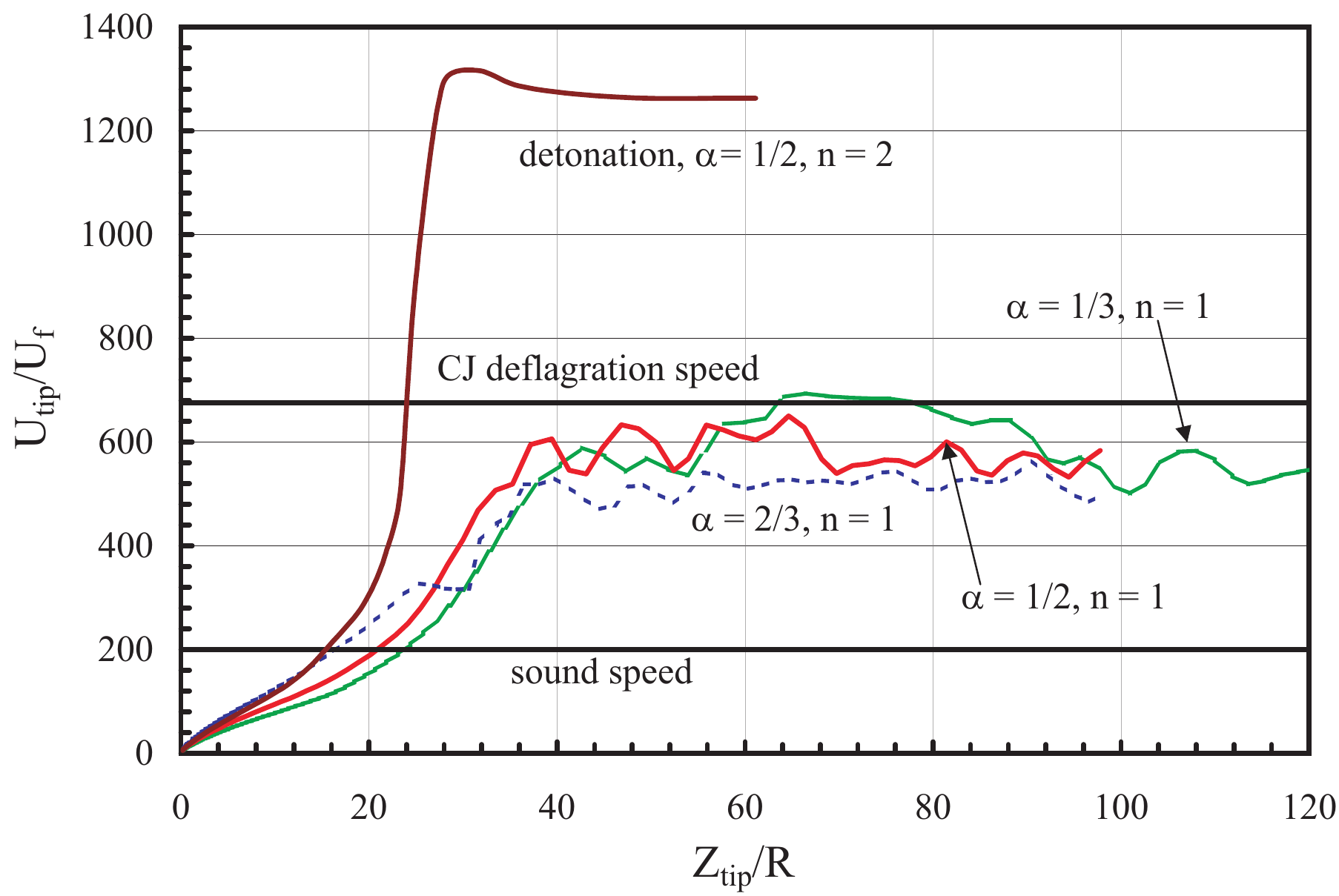}
\caption{ Scaled velocity of the flame tip versus tip position for
the planar geometry and $\alpha = 1 / 3;\,1 / 2;\,2 / 3$,  $\Delta
z / R = 1 / 4$, and different reaction order with respect to
density $n = 1, 2$. The plot with $n=2$ shows also transition to
detonation.}
 \label{fig-10}
\end{figure}
%============================= Figure 10 ===================================%

Simulation of the flame acceleration in channels/tubes with
obstacles was performed at various flow parameters. In particular,
we used various blockage ratios $\alpha = 1 / 3;\;1 / 2;\;2 / 3$,
various spacings between the obstacles, $\Delta z / R = 1 / 4;\;1 /
2;\;1;\;2$, and various initial Mach numbers. Figure \ref{fig-6}
shows the characteristic flame shape and flow velocity obtained at
various conditions. It is found that, in spite of these differences,
all snapshots demonstrate the same basic feature of flame
acceleration in channels with obstacles described theoretically in
Sec.~2, namely, the flame tip propagates fast along the unobstructed
part of the channel, leaving pockets of unburned fuel mixture in
between the obstacles. Delayed burning between the obstacles
involves various levels of turbulence created by the flow. In Fig.
\ref{fig-6} (a) the flow is laminar, and this regime is quite
similar to the theoretical one sketched in Fig. \ref{fig-1} and
described in Sec. 2. In contrast, Fig. \ref{fig-6} (g) shows rather
strong turbulence corresponding to the developed stage of flame
acceleration. Figures \ref{fig-6} (c, e) show the flame shape for
relatively large obstacle spacing: the snapshots are similar to the
experimental photos of accelerating flames in Ref.
\cite{Johansen-Ciccarelli-2007}. We also observe a strong jet-flow
developing in the unobstructed part of the channel in Fig.
\ref{fig-6} (b, d, f, h) which is an important part of the
acceleration mechanism described in Sec. 2. Figure \ref{fig-u-axial}
shows the z-velocity component of the jet along the channel axis for
the initial laminar stages of burning. As predicted by the theory,
Eq. (\ref{eq3}), plotted by the dashed line, we observe almost
linear increase of the gas velocity from the closed end of the tube
to the flame tip. The x-velocity component in Fig. \ref{fig-u-x}
also demonstrates good agreement between the theory and simulation.
Comparison of the theory and simulation in Figs. \ref{fig-u-axial},
\ref{fig-u-x} supports the potential flow model in the burnt gas
employed in the analytical theory of Sec. 2 and of Ref.
\cite{Bychkov-et-al-2008}. The slight deviation between the
theoretical and numerical results is due to the finite flame
thickness and viscosity employed in the simulation. For example,
because of viscous friction, the z-velocity component in the
simulation inevitably decreases at the border between the obstructed
and unobstructed parts of the channel, which produces vorticity in
that region as shown in Fig.~1~(c) of Ref.
\cite{Bychkov-et-al-2008}. Another interesting question is related
to the vorticity generated at the curved flame front as predicted by
the classical theory \cite{Zeldovich.et.al-1985}. In the present
case the curved shape of the flame tip plays a minor role in the
flame dynamics in comparison with the powerful acceleration
mechanism described in Sec.~2. For example, replacing the curved
flame tip by a planar one in Fig. \ref{fig-6}~(a), we obtained
negligible change of the total flame surface area as well as the
total burning rate. Similarly, the vorticity generated by the curved
flame tip plays a minor role in comparison with the strong jet flow
described in Sec.~2 and with the vorticity generated in that flow
because of viscous forces. In this sense the present problem is
completely different from, say, the Darrieus-Landau instability for
which a curved flame shape and the flame velocity increase are
intrinsically related to the vorticity production
\cite{Bychkov-Liberman-2000}.

%============================= Figure 11 ===================================%
\begin{figure}
\includegraphics[width=\columnwidth]{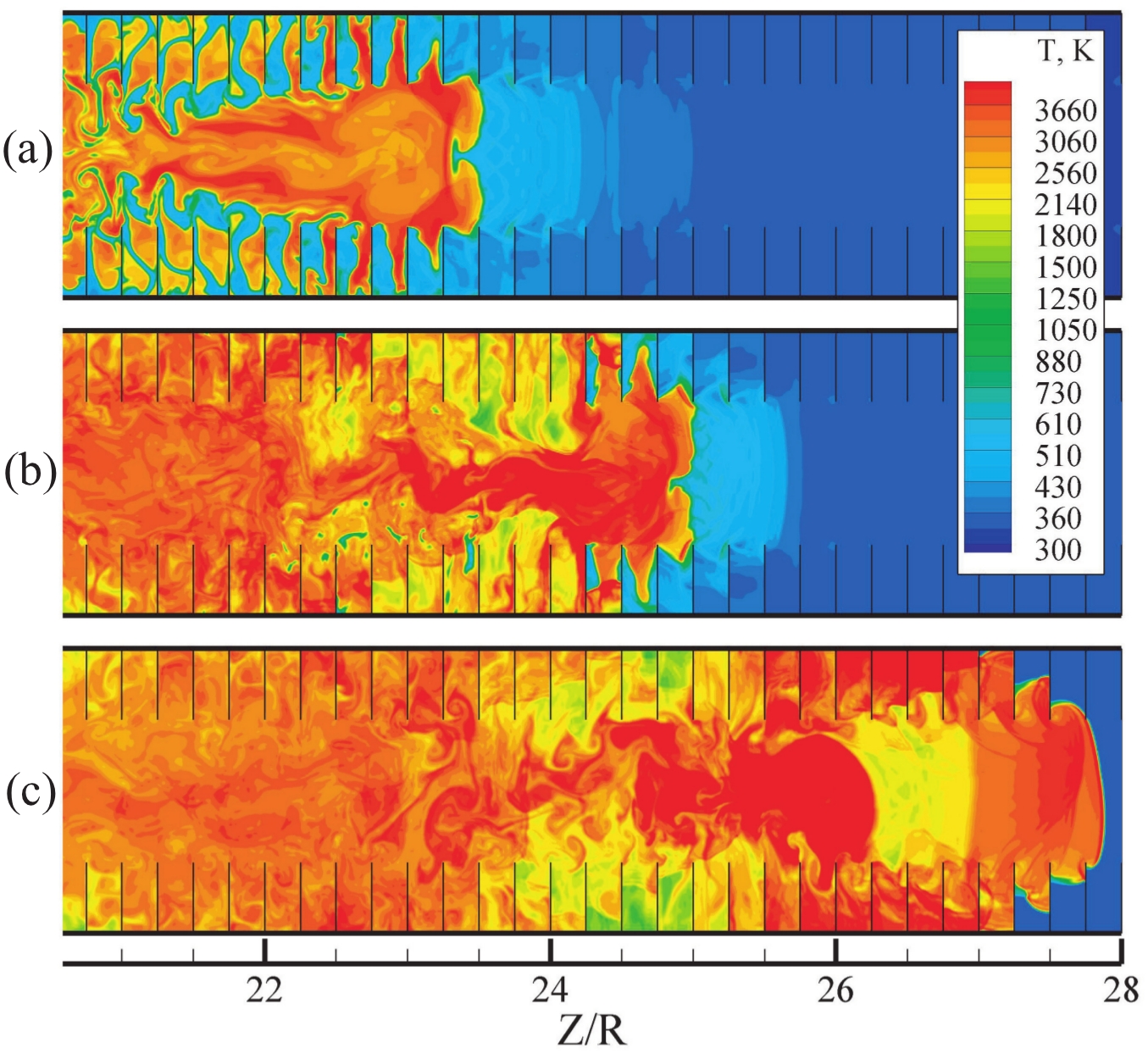}
\caption{ Temperature field for three consecutive moments during
deflagration-to-detonation transition for the second-order reaction
for $\Theta = 8$, $Ma= 0.005$, $\Delta z/R=1/4$, $\alpha = 1/2$.}
\label{fig-11}
\end{figure}
%============================= Figure 11 ===================================%

One of the most interesting questions in this study is to identify
how the flame acceleration depends on the obstacle parameters:
blockage ratio and spacing, i.e. pocket depth and width. According
to the theory of Sec.~2, the acceleration rate is determined  by the
size of the unobstructed part of the channel, $R(1-\alpha)$. In the
theoretical model, the pocket depth and width do not influence the
acceleration, provided that the depth is not too small and the width
is not too large, for which the notion of a pocket becomes
meaningless. Taking proper scaling of the tip position
$Z_{f}/R(1-\alpha)$, we observe that the numerical results reproduce
the theoretical predictions for different obstacle parameters, see
Fig. \ref{fig-7}. Additional numerical data, which is not presented
in Fig. \ref{fig-7} to avoid clutter, can be found in Ref.
\cite{Bychkov-et-al-2008}. In the event of small spacing between the
obstacles ($\Delta z / R = 1 / 4$), the flame tip accelerates
monotonically as described by the theoretical model. When the
spacing is considerable ($\Delta z / R = 2$), we observe strong
turbulent pulsations with space period well-correlated with the
distance between the obstacles. Still, even in this case, the
average velocity of the flame tip is predicted by the theory of Sec.~2 
with good accuracy. Figure \ref{fig-7} also shows good agreement
of the theory and simulation with the experimental results
\cite{Johansen-Ciccarelli-2009}, which also demonstrate noticeable
velocity pulsations because of the large spacing between the
obstacles. When compared to the obstacle positions (not shown in the
figure because of the chosen scaling), both simulation and
experiment demonstrate an increase in the flame tip velocity at
every obstacle. Maximum pulsation velocity corresponds approximately
to the middle between two obstacles.

%============================= Figure 12 ===================================%
\begin{figure}[t]
\includegraphics[width=\columnwidth]{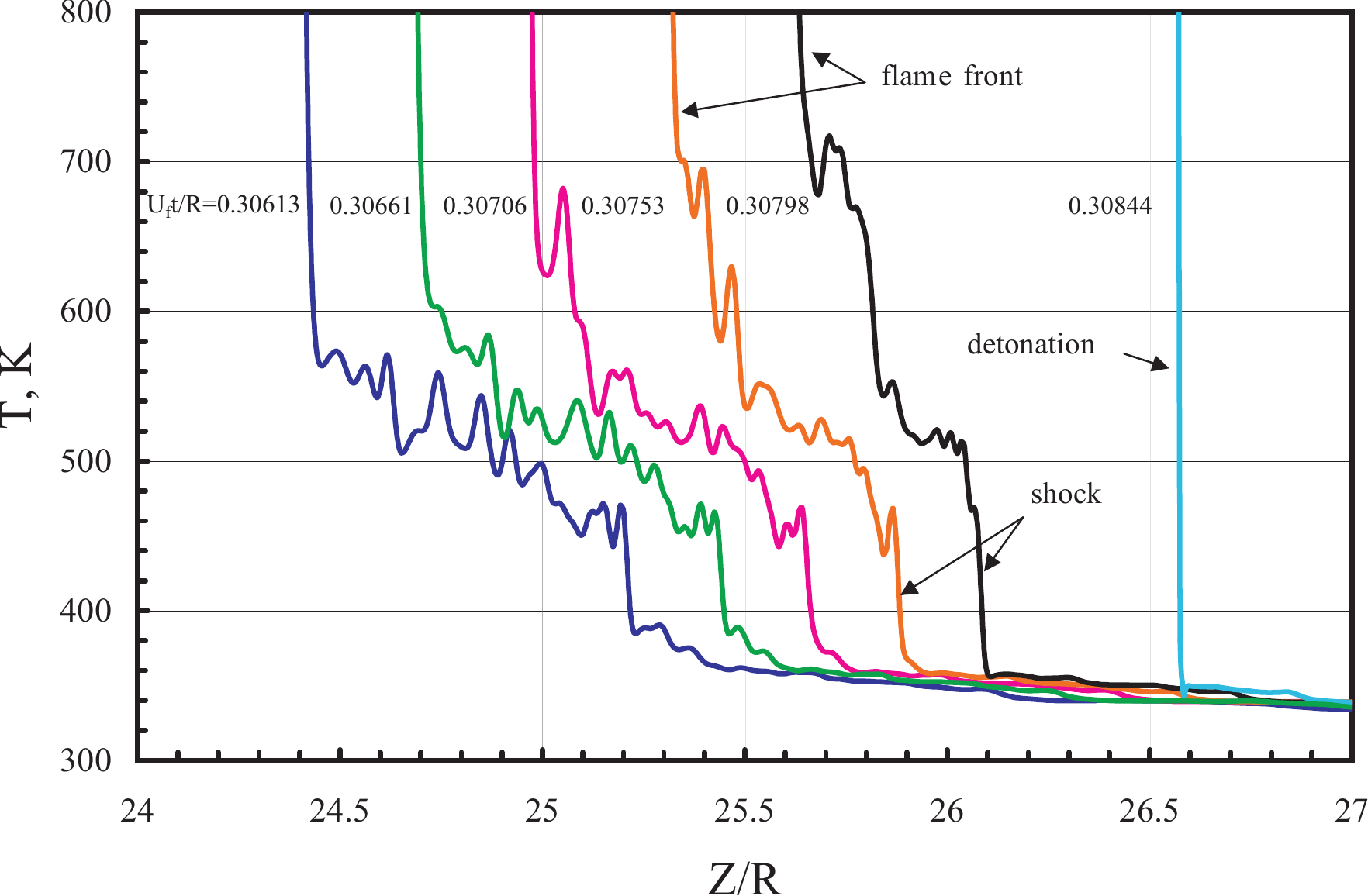}
\caption{ Temperature  profiles along the channel axis at different
time instants close to the instant of detonation triggering; other
parameters are the same as for Fig. \ref{fig-11}. }
 \label{fig-12}
\end{figure}
%============================= Figure 12 ===================================%

For the axisymmetric geometry, the theory of Sec.~2 predicts
considerably stronger flame acceleration as compared to the planar
case. Figure \ref{fig-7} shows the velocity of the flame tip versus
the position in the tube for $Ma = 10^{ - 3}$, $\alpha = 1 / 2;\;2 /
3$, and demonstrates much faster flame acceleration. The growth rate
is about twice larger in agreement with Eqs. (\ref{eq32}),
(\ref{eq34}). However, quantitative agreement between the theory and
simulation in the axisymmetric case is not as good as in the planar
geometry. In the planar case of Fig. \ref{fig-7}, the quantitative
difference is less than 5\%, while in the axisymmetric case this
difference increases from about 10\% at the beginning to 20\% at
later time. Equation (\ref{eq34}) provides better agreement with the
simulation than Eq. (\ref{eq41}), which may indicate a minor role of
the radius growth for the flame skirt in the pockets. This effect
may be observed directly for the flame shape in the axisymmetric
geometry shown in Fig. \ref{fig-8}, which shows that even minor
penetration of the flame skirt into the pockets between the
obstacles provides quite strong flame acceleration in the
axisymmetric geometry. For comparison, the flame skirt penetrates
much deeper into the pockets in the planar case of Fig. \ref{fig-6}~(a).
%\textit{Still, as suggested by one of the referees, fair agreement
%of Eq. (\ref{eq41}) with simulations may be related to other effects
%like compressibility and viscosity.}
We also observe that the difference between theory and modeling
becomes larger in Fig. \ref{fig-7} as the flame velocity increases.
The quantitative difference of 10-20\% in the axisymmetric case is
partly related to the influence of viscosity and the moderate tube
width, corresponding to moderate values of the Reynolds number. As
obtained in numerical studies of the early acceleration of finger
flames \cite{Bychkov-et-al-2007}, moderate tube width reduces flame
acceleration as compared to the theory. However, there are also
other reasons that cause a reduction of the flame acceleration in
the numerical modeling. The first is the incompressibility
assumption of the theory. This assumption is acceptable at the
beginning of the acceleration, but deteriorates as the local Mach
number increases as the flame accelerates. The deviation develops
faster for the axisymmetric geometry because of its faster
acceleration. In order to study the influence of gas compression on
the flame acceleration, we investigated the dependence of the scaled
acceleration rate $\sigma $ on the initial Mach number, as shown in
Fig. \ref{fig-9} for various values of the blockage ratio. It is
thus seen that the scaled acceleration rate decreases strongly with
the Mach number. For initial Mach number $Ma = 10^{-2}$ instead of
$Ma = 10^{-3}$, the acceleration rate is approximately reduced by a
factor of two.

%============================= Figure 13 ===================================%
\begin{figure}[t]
\includegraphics[width=\columnwidth]{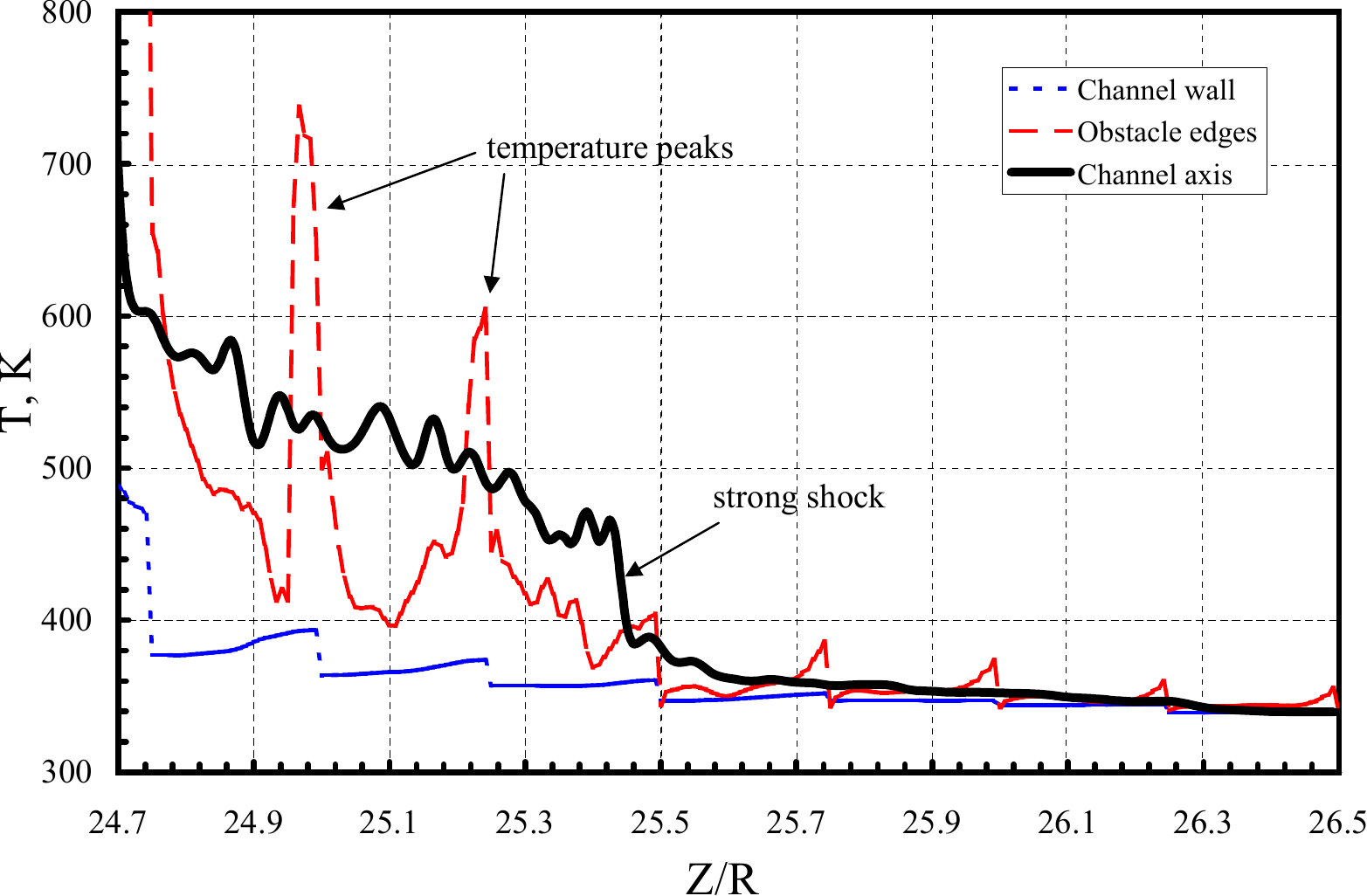}
\caption{ Temperature  profiles along the channel axis, at the
obstacle edges and at the wall for  $U_{f}t/R = 0.30661$; other
parameters are the same as for Fig.  \ref{fig-11}. }
 \label{fig-13}
\end{figure}
%============================= Figure 13 ===================================%

The slowdown of flame acceleration because of gas compression agrees
with the concept that the flame propagation velocity cannot exceed
the value of the Chapman-Jouguet (CJ) limiting state, for which the
downstream flow is sonic. We therefore expect saturation of the
flame tip velocity to a certain steady value at the end of the
acceleration process, but prior to an explosion. Indeed, Fig.
\ref{fig-10} demonstrates for the planar geometry such a saturation,
obtained in the numerical simulation at the final stage of flame
acceleration, for various blockage ratios. The initial Mach number
in Fig. \ref{fig-10} is $Ma = 5 \cdot 10^{-3}$, and the saturation
velocity is about $(2.5 - 3.0)c_{s} $. For comparison,
one-dimensional theory predicts the CJ deflagration velocity with
respect to the wall in the limit of large energy release ($\Theta >
> 1)$ as \cite{Landau&Lifshitz-1989,Chue-et-al-1993}
\begin{equation}
\label{eq59} {\frac{{U_{CJ}}} {{c_{s}}} } = {\left[ {1 +
{\frac{{\gamma (\gamma - 1)}}{{2(\gamma + 1)}}}} \right]}\sqrt
{2{\frac{{\Theta - 1}}{{\gamma + 1}}}}.
\end{equation}
In the present case of $\gamma = 1.4$, $\Theta = 8$, Eq.
(\ref{eq59}) yields $U_{CJ} \approx 2.68c_{s} $, which is about the
saturation velocity obtained in the present numerical simulation.
More accurate theoretical calculation of the CJ deflagration
velocity without the assumption of $\Theta > > 1$ yields $U_{CJ}
\approx 3.38c_{s}$, which is also close to the present numerical
results.

%============================= Figure 14 ===================================%
\begin{figure}
\subfigure[]{\includegraphics[width=0.45\textwidth\centering]{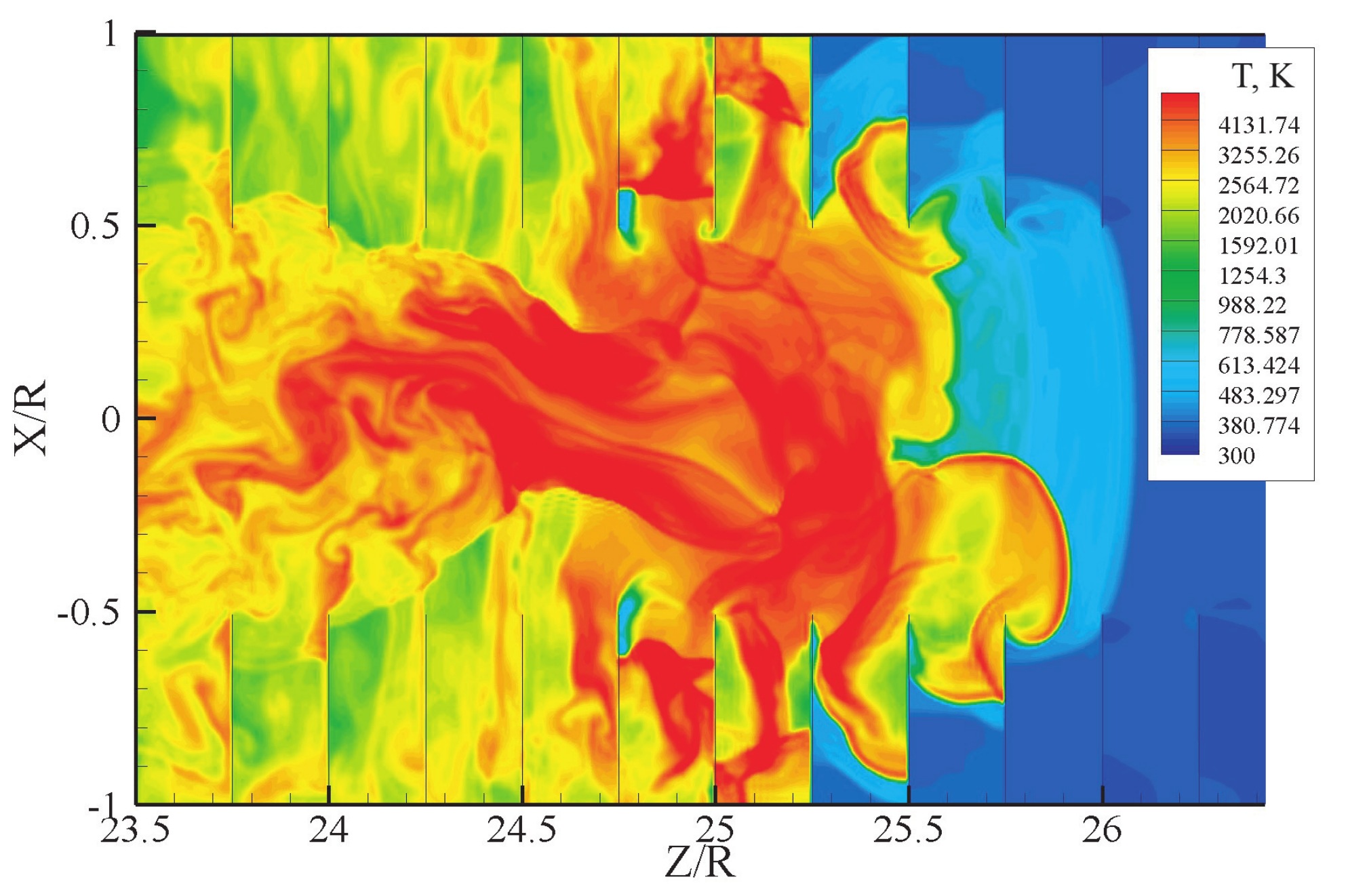}}
\subfigure[]{\includegraphics[width=0.45\textwidth\centering]{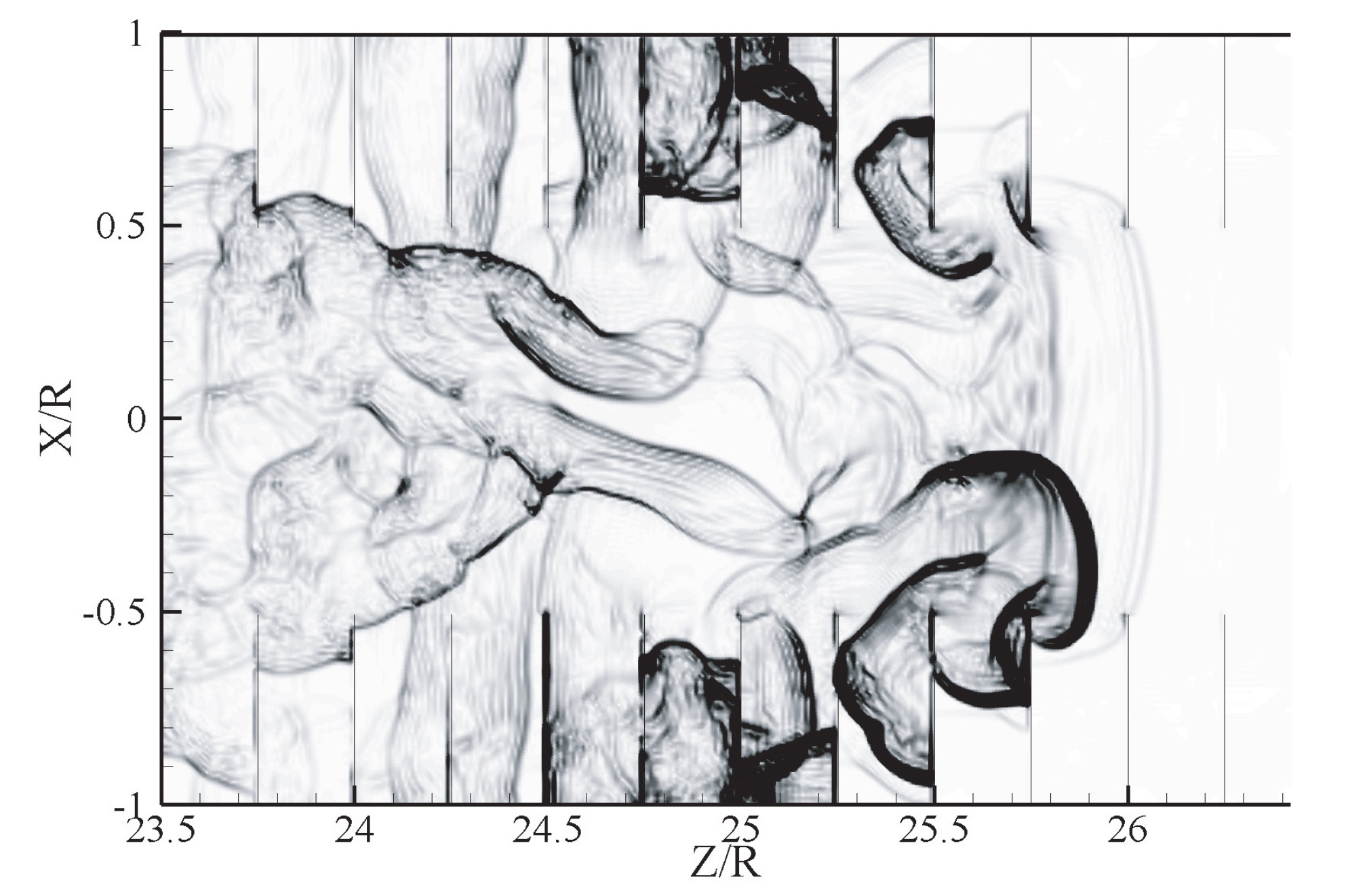}}
\subfigure[]{\includegraphics[width=0.45\textwidth\centering]{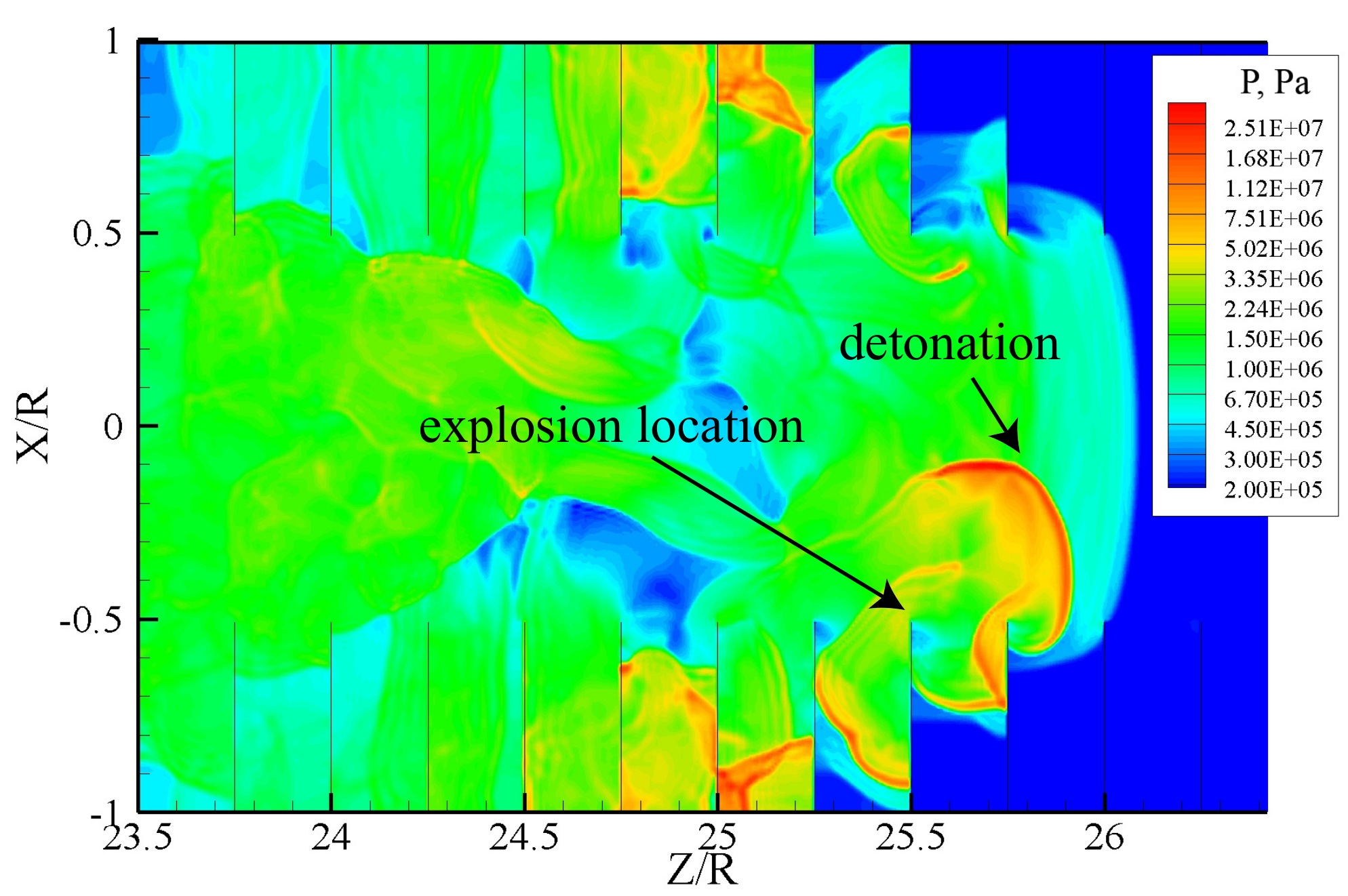}}
\caption{(a) Temperature, (b) pressure gradient modulus and (c)
pressure fields close to the point of explosion triggering for the
second-order reaction for $\Theta = 8$, $Ma= 0.005$, $\Delta
z/R=1/4$, $\alpha = 1/2$.} \label{fig-14}
\end{figure}
%===========================================================================%

Finally, we discuss how flame acceleration in channels with
obstacles may lead to DDT. Since the time and position of DDT are
quite sensitive to  the chemical kinetics adopted, our results on
detonation triggering should be considered as qualitative. At the
same time, they demonstrate the general features of DDT irrespective
of the particular fuel mixture. It is well known that any flame
propagating from a closed end pushes a flow in the fuel mixture with
a weak shock/compression wave at the head of the flow. The flame
acceleration renders the compression wave stronger, until it
develops into a shock of considerable amplitude. Preheating of the
fuel mixture by the shock is conventionally considered as one of the
main elements of DDT both in smooth tubes and in channels with
obstacles \cite{Shelkin,Zeldovich.et.al-1985,Utriev&Oppenheim-1966,
Shepherd.et.al-1992,Roy-et-al-2004,Ciccarelli-Dorofeev-2008}. The
temperature behind the shock increases and the reaction time in any
compressed gas parcel decreases drastically. The decrease in the
reaction time may result in explosion and DDT ahead of the flame
front unless the parcel is burnt by the flame before active
explosion is initiated. Thus, in general, we may expect two possible
outcomes for the flame acceleration: 1) If the reaction time behind
the shock is sufficiently short, then it drives the explosion and
DDT; 2) The reaction time may be longer than the interval available
for a gas parcel to travel between the shock and the flame. In this
case explosion does not occur and the final state of flame
acceleration is the CJ deflagration. For comparison, the possibility
of spontaneous explosion ahead of an accelerating flame was
considered in the theory \cite{Bychkov-Akkerman-2006} for smooth
tubes. Both CJ detonation and deflagration have also been found in
smooth tubes experimentally in Ref. \cite{Wu.et.al-2007}. In
channels with obstacles, the state of CJ deflagration is also known
as "fast flames"
\cite{Ciccarelli-Dorofeev-2008,Kuznetsov-et-al-2005}. In the present
simulations we observed both possibilities of DDT and CJ
deflagration for different reaction kinetics. Taking reaction of the
first order with respect to density (designated by $n=1$ in Fig.
\ref{fig-10}), we obtained statistically steady CJ deflagration at
the end of flame acceleration with no explosion or DDT, see Fig.
\ref{fig-10}. This result indicates that the decrease in the
reaction time behind the shock is not sufficient, and gas parcels
are consumed by the flame front before spontaneous reaction develops
into a powerful explosion. Thus, in order to observe DDT, we need to
take another reaction mechanism, which is more sensitive to pressure
and temperature increase in the shock. Similar to Ref.
\cite{Kagan-et-al-2006}, we considered a reaction of the second
order with respect to density, $n=2$, and obtained explosion
triggering and DDT, see Fig. \ref{fig-10}. Remarkably, in this case
the reaction rate is so sensitive to pressure and temperature that
the DDT occurs before the flame reaches the CJ deflagration state.
Figure \ref{fig-11} shows characteristic temperature snapshots at
the DDT. It is seen that the accelerating flame acts like a piston
pushing a shock, which raises the temperature of the fuel mixture
ahead of the flame, as shown in Fig. \ref{fig-11} (a, b). The
snapshot of Fig. \ref{fig-11} (c) already corresponds to detonation.
Development of the temperature profiles ahead of the flame front
along the channel axis is shown in Fig. \ref{fig-12} for the time
instants $U_{f}t/R = 0.30613 - 0.30844$ close to the DDT time.
Unlike the case of smooth tube with monotonic temperature
distribution in the compression wave \cite{Bychkov-Akkerman-2006},
in Fig. \ref{fig-12} we observe noticeable temperature pulsation due
to secondary shocks reflected from the obstacles. The main tendency
nevertheless remains the same: we can see a considerable temperature
jump in the main shock followed by further temperature increase in
the compression wave from the shock to the flame front. The
compression wave and the shock become stronger as the flame
accelerates, until explosion starts and develops into detonation.

Recent papers \cite{Valiev-et-al-2009,Valiev-et-al-2008} on DDT in
tubes with smooth adiabatic wall have also demonstrated the
important role of viscous heating at the wall in addition to shock
heating. Because of viscous heating, the temperature at the wall of
a smooth adiabatic channel is larger than that at the axis, and
numerical modeling demonstrates DDT onsets at the wall. A similar
physical mechanism of viscous heating may also be identified in
channels with obstacles, though with proper modifications due to the
specific geometry. To elucidate the mechanism, Fig. \ref{fig-13}
shows the temperature profiles for $U_{f}t/R = 0.30661$ along the
channel axis, along the wall and at specific obstacle edges. We can
see that on average, temperature is considerably higher at the
channel axis. Obstacles reduce the shock strength, which results in
much lower average temperature at the obstacle edges, with the
lowest temperature at the wall deep in the pockets. Still, obstacles
not only moderate the main shock, but they also produce hot spots,
which may be crucial in explosion triggering and DDT
\cite{Zeldovich.et.al-1985,Oran-Gamezo-2007}. Though the average
temperature is lower at the obstacle edges, we also observe sharp
peaks of temperature ahead of every obstacle, which are much higher
than the respective temperature at the channel axis. There are two
possible reasons for these temperature peaks. First, secondary shock
waves reflected from the obstacles may produce local temperature
increase. Second, the strong jet-flow pushed by an accelerating
flame slows down at the obstacles and generates vortices in the
pockets with high velocity gradients. Slowdown of the jet flow
increases local pressure and temperature. Viscous dissipation of the
vortices leads to additional temperature increase, which has the
same effect as viscous heating at the wall in smooth tubes
\cite{Valiev-et-al-2009,Valiev-et-al-2008}. The important role of
viscous heating in producing temperature peaks is especially obvious
in Fig. \ref{fig-13} at the positions $Z/R = 25.75; \ 26;\ 26.25$,
which are ahead of the strong shock position, $Z/R = 25.44$. At this
simulation run explosion starts at an obstacle edge, as shown in
Fig. \ref{fig-14}. It is noted the exact position of explosion
triggering may depend on the particular obstacle geometry, e.g. on
the spacing between the obstacles. In Figs. \ref{fig-11},
\ref{fig-14} we used a rather small spacing $\Delta z /R = 1/4$.
Other simulations \cite{Gamezo-et-al-2007} performed for larger
spacings also demonstrated the possibility of explosion triggering
deep in the pockets with the hot spots produced by reflected shocks.
Thus, both in Ref. \cite{Gamezo-et-al-2007} and in the present
simulation, obstacles play an important role in explosion triggering
and DDT.

\section{5. Summary}

This paper presents the theory and numerical simulation of flame
acceleration in channels with obstacles. We showed theoretically as
well as computationally that flame acceleration is noticeably
stronger in the axisymmetric geometry as compared to the planar one.
We also considered the influence of gas compression on the flame
acceleration, and showed numerically that the flame acceleration
rate decreases with increasing initial Mach number, and that the
velocity of the accelerating flame eventually saturates to a value
that is supersonic with respect to the wall and is correlated to the
known CJ deflagration speed. This saturation state has been
referred to as that of fast flames in experimental studies
\cite{Ciccarelli-Dorofeev-2008,Kuznetsov-et-al-2002}. We also
demonstrated numerically the possibility of DDT in the geometry of
obstructed channels.

\section{Aknowledgments}

This work was mostly supported by the Swedish Research Council (VR)
and the Swedish Kempe Foundation. The numerical simulation was
performed at the High Performance Computer Center North (HPC2N),
Umea, Sweden, under SNAC project 007-07-25. The work at Princeton
University was supported by the US Air Force Office of Scientific
Research.

%===================================================================%
%                            References                             %
%===================================================================%
%\newpage

%\newpage

%\centerline{FIGURE CAPTIONS}

\end{document}